\documentclass{article}

\title{Scale-dependent homogeneity measures\\ for causal dynamical triangulations}

\author{Joshua H. Cooperman \\ \emph{Institute for Mathematics, Astrophysics, and Particle Physics}\\ \emph{Radboud Universiteit Nijmegen, Heyendaalseweg 135, 6526 AJ Nijmegen, Nederland}}

\usepackage{float}
\usepackage[small]{caption}
\usepackage{multicol}
\usepackage{subfigure}
\usepackage{tabularx}
\usepackage{booktabs}
\usepackage{graphicx}
\usepackage{amsmath}
\usepackage{amsfonts}
\usepackage{latexsym}
\usepackage{mathrsfs}
\usepackage{geometry}
\usepackage{chngcntr}
\counterwithin{figure}{section}
\counterwithin{table}{section}
\numberwithin{equation}{section}

\begin{document}

\maketitle

\begin{abstract}
I propose two scale-dependent measures of the homogeneity of the quantum geometry determined by an ensemble of causal triangulations. The first measure is volumetric, probing the growth of volume with graph geodesic distance. 
The second measure is spectral, probing the return probability of a random walk with diffusion time. 
Both of these measures, particularly the first, are closely related to those used to assess the homogeneity of our own universe on the basis of galaxy redshift surveys. 
I employ these measures to quantify the quantum spacetime homogeneity as well as the temporal evolution of quantum spatial homogeneity of ensembles of causal triangulations in the well-known physical phase. According to these measures, the quantum spacetime geometry exhibits some degree of inhomogeneity on sufficiently small scales and a high degree of homogeneity on sufficiently large scales. This inhomogeneity appears unrelated to the phenomenon of dynamical dimensional reduction. I also uncover evidence for power-law scaling of both the typical scale on which inhomogeneity occurs and the magnitude of inhomogeneity on this scale with the ensemble average spatial volume of the quantum spatial geometries. 
\end{abstract}

\section{Introduction}\label{introduction}

The standard cosmological model employs a particular Friedmann-Lema\^itre-Robertson-Walker (FLRW) spacetime in its description of our universe's large scale structure. These spacetimes are exactly spatially homogeneous and isotropic. 
While one ultimately justifies this choice of spacetime on the basis of its success in describing our universe, one often invokes the cosmological principle---that no place or direction is privileged at any epoch in our universe---as motivation. Since the assumption of a FLRW spacetime embeds the cosmological principle in the standard cosmological model, tests of this model amount in part to tests of the cosmological principle. One can also attempt to test the cosmological principle in a model-independent  fashion; however, completely disentangling such a test from any model is quite nontrivial \cite{Alonso}. 


Of course, our universe is manifestly inhomogeneous and anisotropic on relatively small length scales, evidently becoming approximately homogeneous and isotropic on relatively large length scales. To account for the inhomogeneity and anisotropy of our universe, 
one introduces coupled gravitational and material perturbations which propagate on the background of the FLRW spacetime. 
These perturbations must maintain magnitudes sufficiently small that their backreaction on the FLRW spacetime is essentially negligible. 
In light of this inhomogeneity and anisotropy on relatively small scales, one actually attempts to determine 
on what relatively large scales homogeneity and isotropy emerge at a particular epoch. Ideally, one would make this determination for every epoch, testing one's prediction for the evolution of these scales, but, currently, 
only certain epochs are accessible to such an analysis. 

For an epoch approximately $3\cdot 10^{5}$ years after the big bang, the cosmic microwave background radiation provides measures of the homogeneity and isotropy of our universe. The magnitude of inhomogeneities at this epoch, quantified by the ratio of the standard deviation to the mean radiation temperature, is approximately $10^{-5}$. In the standard cosmological model these inhomogeneities have their origin in cosmic inflation. 
A model of cosmic inflation predicts the complete spectrum but not the overall magnitude of these inhomogeneities; rather, one first inputs the magnitude, and the model then outputs the spectrum. 
One would of course like an early universe model that predicts the magnitude as well as the spectrum of the inhomogeneities and anisotropies. One often surmises that a quantum theory of gravity would produce such predictions. Various approaches to quantum cosmology inspired by candidate quantum theories of gravity make contact with the standard cosmological model and make predictions for early universe cosmology \cite{QGQC}. As far as I know, however, none of these approaches can claim to give a first principles derivation of the magnitudes of inhomogeneity and anisotropy. One should now be seriously working to extract such predictions: although the BICEP2 experiment's detection of a signature of primordial gravitational perturbations is in doubt \cite{BICEP2,RF&JCH&DNS}, more robust findings are likely just over the horizon.

To investigate the cosmological principle at much more recent epochs, one studies the distribution of matter in our universe through galaxy redshifts surveys. 
The number of galaxies within a $3$-sphere of a given physical radius at a fixed redshift serves as a standard quantity for ascertaining the scale on which homogeneity emerges \cite{SDSS}. 
One varies the radius to find the scaling of this number of galaxies with the radius. For a homogeneous distribution of particles in $3$-dimensional Euclidean space, the number of particles within a $3$-sphere scales with its radius cubed. Since the standard cosmological model takes space at a fixed redshift as $3$-dimensional Euclidean space, scaling of the number of galaxies with the radius cubed constitutes evidence for homogeneity. 
One thus estimates the scale on which homogeneity emerges as the radius at which this scaling first appears. Relatedly, the power with which each higher moment of the number of galaxies within a $3$-sphere scales with the radius---its so-called fractal dimension---also serves as a measure of homogeneity \cite{WiggleZ}.

As an extremely preliminary foray in the direction of predicting the magnitude of inhomogeneity from a quantum theory of gravity, I propose two scale-dependent homogeneity measures 
for causal dynamical triangulations. After briefly introducing this approach to the construction of quantum theories of gravity in section \ref{background}, I formally define the two homogeneity measures in section \ref{definitions}. Both of these measures draw inspiration from the above techniques for assessing the homogeneity of our own universe on the basis of galaxy redshift surveys. The first---a volumetric measure---captures the variance in a sphere's volume depending on its central point 
as a function of its radius. This homogeneity measure is closely related to the Hausdorff dimension. The second---a spectral measure---captures the variance in the return probability of a random walk depending on its 
starting point as a function of the number of its steps. This homogeneity measure is closely related to the spectral dimension. 

I report numerical measurements of the two homogeneity measures  in section \ref{results}. I study exclusively the phase of Wick-rotated quantum geometry known to possess semiclassical properties on sufficiently large scales. In particular, this phase's quantum geometry is well described as that of Euclidean de Sitter space on its largest scales \cite{JA&AG&JJ&RL1,JA&AG&JJ&RL2,JA&AG&JJ&RL&JGS&TT,JA&JJ&RL1,JA&JJ&RL2,JA&JJ&RL3,JA&JJ&RL4,JA&JJ&RL5,JA&JJ&RL6,CA&SJC&JHC&PH&RKK&PZ,JHC&KL&JMM,JHC&JMM}. Since the spacetime geometry of our universe is well described as that of Lorentzian de Sitter spacetime both during cosmic inflation and in the far future, examining this phase of quantum geometry seems a not so unreasonable starting point. I apply both homogeneity measures in two capacities: quantifying the homogeneity of the quantum spacetime geometry and quantifying the homogeneity of the distinguished quantum spatial geometries. By tracing the temporal evolution of these quantum spatial geometries, I analyze the epoch dependence of homogeneity in complete analogy to analyses of our universe. 




Applied to the quantum spacetime geometry, both homogeneity measures provide evidence for some degree of inhomogeneity on sufficiently small scales and a high degree of homogeneity on sufficiently large scales. The latter finding is consistent with the aforementioned semiclassical properties of the studied phase of quantum geometry. 
Finite size scaling analyses indicate that the volumetric homogeneity measure, which finite size scales canonically, 
may not be well-defined in the continuum limit 
but that the spectral homogeneity measure, which finite size scales anomalously, may be well-defined in the continuum limit. Applied to the distinguished quantum spatial geometries, both homogeneity measures continue to provide evidence for some degree of inhomogeneity on sufficiently small scales and a high degree of homogeneity on sufficiently large scales. 
According to the volumetric homogeneity measure, the typical scale on which inhomogeneity occurs  exhibits power-law scaling with the ensemble average spatial volumes of the distinguished quantum spatial geometries. According to both homogeneity measures, the typical magnitude of inhomogeneity exhibits subleading power-law scaling with the ensemble average spatial volumes of the distinguished quantum spatial geometries.
I discuss the import of all of these results in section \ref{conclusion}.

\section{Background}\label{background}

Causal dynamical triangulations is an approach to the quantization of classical theories of gravity based on a particular lattice regularization of the corresponding path integral. I introduce the formalism of causal dynamical triangulations in subsection \ref{theory}, the numerical techniques employed in studying causal dynamical triangulations in subsection \ref{numerics}, and the phenomenology of causal dynamical triangulations in subsection \ref{phenomenology}. See \cite{JA&AG&JJ&RL3} for a comprehensive review. 

\subsection{Formalism}\label{theory}

Suppose that one wishes to quantize a classical theory of gravity defined by the action $S_{\mathrm{cl}}[\mathbf{g}]$ as a functional of the spacetime metric tensor $\mathbf{g}$. Employing path integral techniques to define such a quantum theory, one computes its transition amplitudes as
\begin{equation}\label{transamp}
\mathscr{A}[\gamma]=\int_{\mathbf{g}|_{\partial\mathscr{M}}=\gamma}\mathrm{d}\mu(\mathbf{g})\,e^{iS_{\mathrm{cl}}[\mathbf{g}]/\hbar}
\end{equation}
and its associated expectation values of physical observables as
\begin{equation}\label{expvalue}
\mathbb{E}_{\mathscr{A}[\gamma]}[\mathscr{O}]=\int_{\mathbf{g}|_{\partial\mathscr{M}}=\gamma}\mathrm{d}\mu(\mathbf{g})\,e^{iS_{\mathrm{cl}}[\mathbf{g}]/\hbar}\,\mathscr{O}[\mathbf{g}].
\end{equation}
The path integrations in equations \eqref{transamp} and \eqref{expvalue} provide formal instructions for computing the transition amplitudes $\mathscr{A}[\gamma]$ and the expectation values $\mathbb{E}_{\mathscr{A}[\gamma]}[\mathscr{O}]$: integrate over all physically distinct metric tensors $\mathbf{g}$ satisfying the boundary condition $\mathbf{g}|_{\partial\mathscr{M}}=\gamma$, weighting each metric tensor $\mathbf{g}$ by the product of the measure $\mathrm{d}\mu(\mathbf{g})$ and the exponential $e^{iS_{\mathrm{cl}}[\mathbf{g}]/\hbar}$. $\mathscr{M}$ denotes the Lorentzian manifold on which the metric tensor $\mathbf{g}$ is defined. Carrying out these formal instructions is no simple matter.

The causal dynamical triangulations approach postulates a prescription for making these instructions concrete \cite{JA&JJ&RL1,JA&JJ&RL2}. One first hypothesizes that the only physically relevant transition amplitudes (and the associated expectation values of physical observables) are those for which every metric tensor $\mathbf{g}$ is defined on a manifold $\mathscr{M}$ of the form $\Sigma\times\mathcal{I}$, the direct product of a fixed spatial manifold $\Sigma$ and a real temporal interval $\mathcal{I}$. This is the key hypothesis of the causal dynamical triangulations approach; see, for instance, \cite{JA&AG&JJ&RL3} for its motivation. Accordingly, one defines a distinct quantum theory for each choice of the spatial manifold $\Sigma$. One next introduces a particular lattice regularization of the path integrations in equations \eqref{transamp} and \eqref{expvalue} compatible with this manifold structure. Specifically, one considers the transition amplitudes 
\begin{equation}\label{CDTpathsum}
\mathcal{A}_{\Sigma}[\Gamma]=\sum_{\substack{\mathcal{T}_{c} \\ \mathcal{T}_{c}\cong\Sigma\times\mathcal{I} \\ \mathcal{T}_{c}|_{\partial\mathcal{T}_{c}}=\Gamma}}\mu(\mathcal{T}_{c})\,e^{i\mathcal{S}_{\mathrm{cl}}[\mathcal{T}_{c}]/\hbar}.
\end{equation}
and the associated expectation values of discrete observables
\begin{equation}\label{CDTexpvalue}
\mathbb{E}_{\mathcal{A}_{\Sigma}[\Gamma]}[\mathcal{O}]=\sum_{\substack{\mathcal{T}_{c} \\ \mathcal{T}_{c}\cong\Sigma\times\mathcal{I} \\ \mathcal{T}_{c}|_{\partial\mathcal{T}_{c}}=\Gamma}}\mu(\mathcal{T}_{c})\,e^{i\mathcal{S}_{\mathrm{cl}}[\mathcal{T}_{c}]/\hbar}\,\mathcal{O}[\mathcal{T}_{c}].
\end{equation}
The path summations in equations \eqref{CDTpathsum} and \eqref{CDTexpvalue} provide concrete instructions for computing the transition amplitudes $\mathcal{A}_{\Sigma}[\Gamma]$ and the expectation values $\mathbb{E}_{\mathcal{A}_{\Sigma}[\Gamma]}[\mathcal{O}]$: sum over all physically distinct causal triangulations $\mathcal{T}_{c}$ satisfying the boundary condition $\mathcal{T}_{c}|_{\partial\mathcal{T}_{c}}=\Gamma$, weighting each causal triangulation by the product of the measure $\mu(\mathcal{T}_{c})$ and the exponential $e^{i\mathcal{S}_{\mathrm{cl}}[\mathcal{T}_{c}]/\hbar}$. One takes the measure $\mu(\mathcal{T}_{c})$ as the inverse of the order of the automorphism group of the causal triangulation $\mathcal{T}_{c}$ and the action $\mathcal{S}_{\mathrm{cl}}[\mathcal{T}_{c}]$ as the discretization of the action $S_{\mathrm{cl}}[\mathbf{g}]$ in the Regge calculus of causal triangulations.

\begin{figure}[h!]
\centering
\includegraphics[scale=0.35]{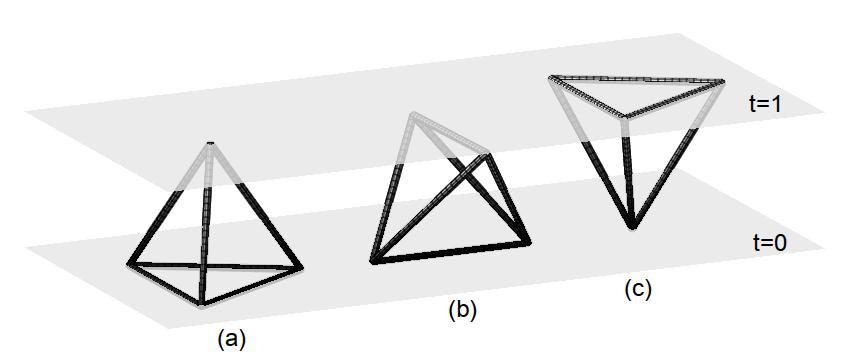}
\caption{Three types of causal $3$-simplices employed in $(2+1)$-dimensional causal dynamical triangulations: (a) $(3,1)$ $3$-simplex, (b) $(2,2)$ $3$-simplex, (c) $(1,3)$ $3$-simplex. The first number in the ordered pair indicates the number of vertices on the initial spacelike $3$-surface ($t=0$), and the second number in the ordered pair indicates the number of vertices on the final spacelike $3$-surface ($t=1$). I denote by $N_{3}^{(3,1)}$ the number of $(3,1)$ $3$-simplices, by $N_{3}^{(2,2)}$ the number of $(2,2)$ $3$-simplices, and by $N_{3}^{(1,3)}$ the number of $(1,3)$ $3$-simplices in a causal triangulation. I have taken this figure from \cite{JHC&JMM}.}
\label{3simplices}
\end{figure}

A $(d+1)$-dimensional causal triangulation $\mathcal{T}_{c}$ is a piecewise-Minkowski simplicial manifold admitting a global foliation by spacelike $d$-surfaces all isomorphic to the spatial manifold $\Sigma$. One constructs a causal triangulation $\mathcal{T}_{c}$ by appropriately joining together $N_{d+1}$ causal $(d+1)$-simplices. A causal $(d+1)$-simplex is a timelike simplicial piece of $(d+1)$-dimensional Minkowski spacetime. 
I depict in figure \ref{3simplices} the three types of causal $3$-simplices. 
Their spacelike edges have squared proper length $a^{2}$, and their timelike edges have squared proper length $-\alpha a^{2}$. The parameter $a$ is the lattice spacing, and the parameter $\alpha$ is a positive real number. 
The $N_{d+1}$ causal $(d+1)$-simplices assemble so as to form spacelike $d$-surfaces, triangulated by regular spacelike $d$-simplices, isomorphic to $\Sigma$, connected by timelike edges. A causal triangulation thus possesses a distinguished foliation by spacelike $d$-surfaces---that distinguished by the skeleton of the causal triangulation. I enumerate the $T$ leaves of the distinguished foliation with a discrete time coordinate $\tau$. 

Since one introduced the path summations \eqref{CDTpathsum} and \eqref{CDTexpvalue} as regularized versions of the path integrations \eqref{transamp} and \eqref{expvalue} (for fixed $\Sigma$), one finally attempts to remove the regularization through a process of renormalization, thereby obtaining the continuum limit of the quantum theory so defined. This process involves attempting to let the lattice spacing $a$ decrease to zero and the number $N_{d+1}$ increase without bound by tuning the couplings of the action $\mathcal{S}_{\mathrm{cl}}[\mathcal{T}_{c}]$ all in such a manner than physical quantities remain finite.

\subsection{Numerics}\label{numerics}

One does not currently know how to evaluate analytically the path summations in equations \eqref{CDTpathsum} and \eqref{CDTexpvalue} (excepting the simplest few cases in $1+1$ dimensions). One thus turns to numerical techniques, in particular, Markov chain Monte Carlo simulations. 
Since such simulations require that the causal triangulations be weighted by real as opposed to complex numbers, one applies a Wick rotation 
consisting of the analytic continuation of the parameter $\alpha$ to $-\alpha$ in the lower half complex plane. This Wick rotation transforms the path summation \eqref{CDTpathsum} into the partition function
\begin{equation}\label{CDTfullpartfunc}
\mathcal{Z}_{\Sigma}[\Gamma]=\sum_{\substack{\mathcal{T}_{c} \\ \mathcal{T}_{c}\cong\Sigma\times\mathcal{I} \\ \mathcal{T}_{c}|_{\partial\mathcal{T}_{c}}=\Gamma}}\mu(\mathcal{T}_{c})\,e^{-\mathcal{S}_{\mathrm{cl}}^{(\mathrm{E})}[\mathcal{T}_{c}]/\hbar}
\end{equation}
for the real-valued Euclidean action $\mathcal{S}_{\mathrm{cl}}^{(\mathrm{E})}[\mathcal{T}_{c}]$. Since such simulations also require that the causal triangulations be finite, one chooses to fix the number $T$ of the distinguished foliation's leaves and the number $N_{d+1}$ of causal $(d+1)$-simplices. This choice leads to the partition function
\begin{equation}\label{CDTpartfunc}
Z_{\Sigma}[\Gamma]=\sum_{\substack{\mathcal{T}_{c} \\ \mathcal{T}_{c}\cong\Sigma\times\mathcal{I} \\ \mathcal{T}_{c}|_{\partial\mathcal{T}_{c}}=\Gamma \\ T(\mathcal{T}_{c})=\bar{T} \\ N_{d+1}(\mathcal{T}_{c})=\bar{N}_{d+1}}}\mu(\mathcal{T}_{c})\,e^{-\mathcal{S}_{\mathrm{cl}}^{(\mathrm{E})}[\mathcal{T}_{c}]/\hbar},
\end{equation}
related to the partition function \eqref{CDTfullpartfunc} by a Legendre transform. One then runs Markov chain Monte Carlo simulations of causal triangulations representative of those contributing to the partition function \eqref{CDTpartfunc}. 

In the following I take the action $S_{\mathrm{cl}}[\mathbf{g}]$ to be the $(2+1)$-dimensional Einstein-Hilbert action
\begin{equation}
S_{\mathrm{cl}}[\mathbf{g}]=\frac{1}{16\pi G}\int_{\mathscr{M}}\mathrm{d}^{3}x\sqrt{-g}(R-2\Lambda)
\end{equation} 
for positive cosmological constant $\Lambda$, and I exclusively consider the case in which each of the distinguished foliation's leaves 
has the topology of a $2$-sphere $\mathrm{S}^{2}$ and the discrete time coordinate $\tau$ has the topology of a $1$-sphere $\mathrm{S}^{1}$. Ambj\o rn \emph{et al} derived the corresponding action $\mathcal{S}_{\mathrm{cl}}^{(\mathrm{E})}[\mathcal{T}_{c}]$ \cite{JA&JJ&RL2}:
\begin{equation}\label{ECDTaction3}
\mathcal{S}_{\mathrm{cl}}^{(\mathrm{E})}[\mathcal{T}_{c}]=-k_{0}N_{0}+k_{3}N_{3}. 
\end{equation}
The bare couplings $k_{0}$ and $k_{3}$ are particular functions of $Ga^{-1}$ and $\Lambda a^{2}$, $N_{0}$ is the number of vertices, and the specific value of the parameter $\alpha$ is irrelevant for $d=2$. 
For particular fixed values of the number $T$, the number $N_{3}$, and the bare coupling $k_{0}$,\footnote{Given these fixed values, one must tune the bare coupling $k_{3}$ to criticality to render well-defined the partition function \eqref{CDTpartfunc} for the action \eqref{ECDTaction3}.} one's Markov chain Monte Carlo simulation generates an ensemble of $N(\mathcal{T}_{c})$ causal triangulations representative of those contributing to the partition function \eqref{CDTpartfunc} for the action \eqref{ECDTaction3}. Given a discrete observable $\mathcal{O}$, one estimates its expectation value 
\begin{equation}
\mathbb{E}_{Z_{\Sigma}[\Gamma]}[\mathcal{O}]=\sum_{\substack{\mathcal{T}_{c} \\ \mathcal{T}_{c}\cong\Sigma\times\mathcal{I} \\ \mathcal{T}_{c}|_{\partial\mathcal{T}_{c}}=\Gamma \\ T(\mathcal{T}_{c})=\bar{T} \\ N_{d+1}(\mathcal{T}_{c})=\bar{N}_{d+1}}}\mu(\mathcal{T}_{c})\,e^{-\mathcal{S}_{\mathrm{cl}}^{(\mathrm{E})}[\mathcal{T}_{c}]/\hbar}\mathcal{O}[\mathcal{T}_{c}]
\end{equation}
in the quantum state specified by the partition function \eqref{CDTpartfunc} as the average 
\begin{equation}
\langle\mathcal{O}\rangle=\frac{1}{N(\mathcal{T}_{c})}\sum_{l=1}^{N(\mathcal{T}_{c})}\mathcal{O}[\mathcal{T}_{c}^{(l)}]
\end{equation}
over the ensemble of $N(\mathcal{T}_{c})$ causal triangulations. In the limit as the number $N(\mathcal{T}_{c})$ of causal triangulations increases without bound, the Metropolis algorithm behind the Markov chain Monte Carlo simulations guarantees that the ensemble average $\langle\mathcal{O}\rangle$ converges to the expectation value $\mathbb{E}_{Z_{\Sigma}[\Gamma]}[\mathcal{O}]$. 

\subsection{Phenomenology}\label{phenomenology}

The partition function \eqref{CDTpartfunc} for the action \eqref{ECDTaction3} exhibits two phases of quantum geometry, the decoupled phase A and the physical phase C, separated by a first order transition \cite{JA&JJ&RL3,RK}. In the following I consider exclusively phase C, which possesses semiclassical characteristics on sufficiently large scales \cite{JA&JJ&RL3,CA&SJC&JHC&PH&RKK&PZ,DB&JH,JHC&KL&JMM,JHC&JMM,RK}. 
This phase structure differs from that of the corresponding partition function for the physically relevant case of the $(3+1)$-dimensional Einstein-Hilbert action, specifically, in lacking a crumpled phase B and the second order transition between phases B and C \cite{JA&SJ&JJ&RL1,JA&SJ&JJ&RL2}. The physical properties of phase C in the two cases are, however, equivalent in all known respects. This statement applies to the interior of phase C, where the majority of studies have been performed, not to the boundaries of phase C, where the difference in phase structure becomes manifest. I have thus chosen to investigate the $(2+1)$-dimensional quantum theory so that the required computing time is significantly less.



\begin{figure}[h!]
\centering
\subfigure[ ]{
\includegraphics[scale=0.95]{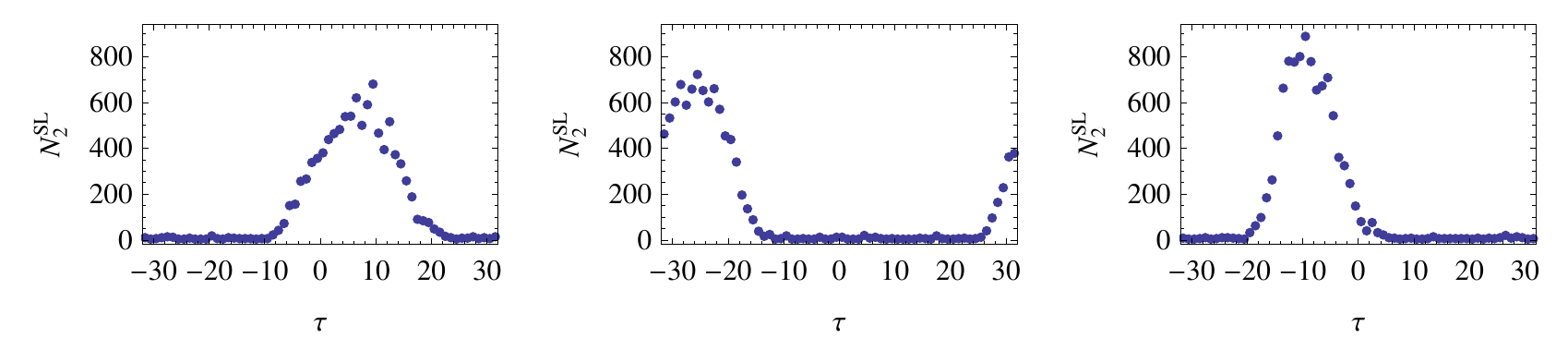}
\label{raw}
}
\subfigure[ ]{
\includegraphics[scale=0.95]{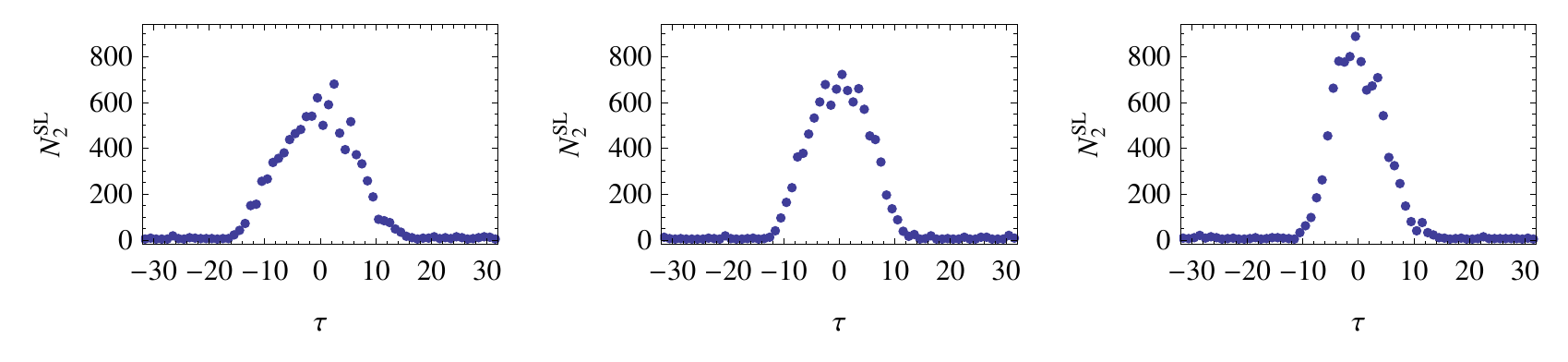}
\label{aligned}
}
\caption[Optional caption for list of figures]{\subref{raw} Number $N_{2}^{\mathrm{SL}}$ of spacelike $2$-simplices as a function of the discrete time coordinate $\tau$ for three causal triangulations from an ensemble characterized by $T=64$, $N_{3}=30850$, and $k_{0}=1.0$. \subref{aligned} Number $N_{2}^{\mathrm{SL}}$ of spacelike $2$-simplices as a function of the  shifted discrete time coordinate $\tau$ for three causal triangulations from an ensemble characterized by $T=64$, $N_{3}=30850$, and $k_{0}=1.0$.}
\label{coherentaveraging}
\end{figure}

Consider now the three causal triangulations representative of those contributing to the partition function \eqref{CDTpartfunc} for the action \eqref{ECDTaction3} depicted in figure \ref{coherentaveraging}\subref{raw}. 
Each of these depictions shows $N_{2}^{\mathrm{SL}}(\tau)$, the number $N_{2}^{\mathrm{SL}}$ of spacelike $2$-simplices comprising a leaf of the distinguished foliation as a function of the discrete time coordinate $\tau$. 
One may conceive of the function $N_{2}^{\mathrm{SL}}(\tau)$ as the discrete time evolution of the discrete spatial $2$-volume of a causal triangulation. The function $N_{2}^{\mathrm{SL}}(\tau)$ clearly distinguishes amongst (most of) the distinguished foliation's leaves: aside from those leaves having nearly zero discrete spatial $2$-volume---the so-called stalk---the function $N_{2}^{\mathrm{SL}}(\tau)$ exhibits a modulation from small to large to small values---the so-called central accumulation. The temporal center of discrete spatial $2$-volume falls within the central accumulation of a causal triangulation. By appropriately shifting the discrete time coordinate of a causal triangulation, one can align the former's zero with the latter's temporal center of discrete spatial $2$-volume as depicted in figure \ref{coherentaveraging}\subref{aligned}. Once one has so relabeled the discrete time coordinate for all of the causal triangulations in an ensemble, one can identify spatial geometries across these causal triangulations by their discrete time coordinate values. 



\begin{figure}[h!]
\centering
\subfigure[ ]{
\includegraphics[scale=1]{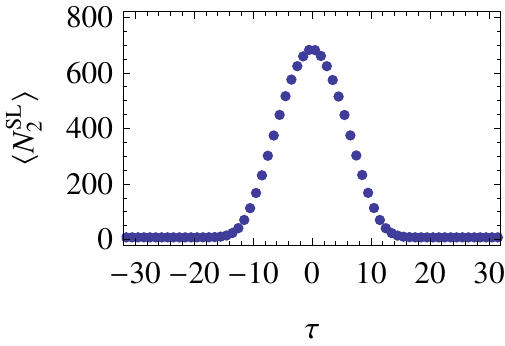}
\label{cohensavvolprof}
}
\subfigure[ ]{
\includegraphics[scale=1]{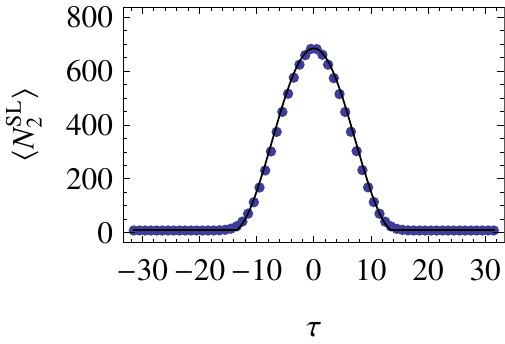}
\label{cohensavvolproffit}
}
\subfigure[ ]{
\includegraphics[scale=1]{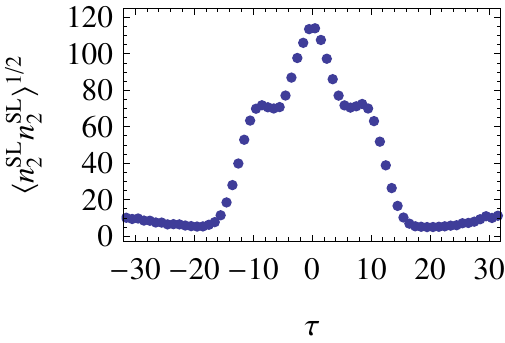}
\label{cohensavvolproffluc}
}
\caption[Optional caption for list of figures]{\subref{cohensavvolprof} Ensemble average number $\langle N_{2}^{\mathrm{SL}}\rangle$ of spacelike $2$-simplices as a function of the discrete time coordinate $\tau$ for an ensemble characterized by $T=64$, $N_{3}=30850$, and $k_{0}=1.0$. \subref{cohensavvolproffit} Fit of the discretization \eqref{discretevolprof} to the ensemble average number $\langle N_{2}^{\mathrm{SL}}\rangle$ of spacelike $2$-simplices as a function of the discrete time coordinate $\tau$ for an ensemble characterized by $T=64$, $N_{3}=30850$, and $k_{0}=1.0$. \subref{cohensavvolproffluc} Diagonal $\langle n_{2}^{\mathrm{SL}}(\tau)n_{2}^{\mathrm{SL}}(\tau)\rangle^{1/2}$ of the ensemble average covariance of deviations $n_{2}^{\mathrm{SL}}(\tau)$ from the ensemble average $\langle N_{2}^{\mathrm{SL}}(\tau)\rangle$ for an ensemble characterized by $T=64$, $N_{3}=30850$, and $k_{0}=1.0$.}
\label{volprof}
\end{figure}

One now studies $\langle N_{2}^{\mathrm{SL}}(\tau)\rangle$, the ensemble average number $\langle N_{2}^{\mathrm{SL}}\rangle$ of spacelike $2$-simplices as a function of the (shifted) discrete time coordinate $\tau$, depicted in figure \ref{volprof}\subref{cohensavvolprof}. 
Restricted to the central accumulation, the ensemble average $\langle N_{2}^{\mathrm{SL}}(\tau)\rangle$ is well described by the discretization
\begin{equation}\label{discretevolprof}
\langle N_{2}^{\mathrm{SL}}(\tau)\rangle=\frac{N_{3}^{2/3}}{\pi s_{0}(1+\xi)}\cos^{2}{\left(\frac{\tau}{s_{0}N_{3}^{1/3}}\right)}
\end{equation}
of the spatial $2$-volume as a function of the global time coordinate of Euclidean de Sitter space \cite{JA&JJ&RL3,CA&SJC&JHC&PH&RKK&PZ,JHC&KL&JMM,JHC&JMM,RK}. $s_{0}$ is the single fit parameter, and $\xi$ is the ratio of the number $N_{3}^{(2,2)}$ of $(2,2)$ $3$-simplices to the combined number $N_{3}^{(1,3)}+N_{3}^{(3,1)}$ of $(1,3)$ and $(3,1)$ $3$-simplices. I show in figure \ref{volprof}\subref{cohensavvolproffit} the fit to the ensemble average $\langle N_{2}^{\mathrm{SL}}(\tau)\rangle$ of the discretization \eqref{discretevolprof}. To arrive at the discretization \eqref{discretevolprof}, one invokes a finite size scaling \emph{Ansatz} relating discrete quantities to their continuous counterparts. In this case one employs the canonical finite size scaling \emph{Ansatz}
\begin{equation}\label{FSSansatz}
 V_{3}=\lim_{\substack{a\rightarrow0 \\ N_{3}\rightarrow\infty}}C_{3}N_{3}a^{3}
\end{equation}
relating the discrete spacetime $3$-volume $N_{3}$ to the continuous spacetime $3$-volume $V_{3}$. $C_{3}$ is the ensemble average effective discrete spacetime $3$-volume of one $3$-simplex.  See, for instance, \cite{JA&JJ&RL6} for justification of the finite size scaling \emph{Ansatz} \eqref{FSSansatz}. On the basis of the \emph{Ansatz} \eqref{FSSansatz}, one scales discrete quantities associated with the units $a^{p}$ by $N_{3}^{-p/3}$. The discrete time coordinate $\tau$ is associated with units of $a$; accordingly, it appears in equation \eqref{discretevolprof} scaled by $N_{3}^{-1/3}$. The ensemble average $\langle N_{2}^{\mathrm{SL}}(\tau)\rangle$ is associated with the units of $a^{2}$; accordingly, it appears in equation \eqref{discretevolprof} scaled by $N_{3}^{-2/3}$ (appearing on the opposite side of this equation).

One also studies $\langle n_{2}^{\mathrm{SL}}(\tau)n_{2}^{\mathrm{SL}}(\tau')\rangle$, the ensemble average covariance of deviations $n_{2}^{\mathrm{SL}}(\tau)$ from the ensemble average $\langle N_{2}^{\mathrm{SL}}(\tau)\rangle$, defined as
\begin{equation}
\langle n_{2}^{\mathrm{SL}}(\tau)n_{2}^{\mathrm{SL}}(\tau')\rangle=\frac{1}{N(\mathcal{T}_{c})}\sum_{l=1}^{N(\mathcal{T}_{c})}\left\{\left[N_{2}^{\mathrm{SL}}(\tau)\right]^{(l)}-\langle N_{2}^{\mathrm{SL}}(\tau)\rangle\right\}\left\{\left[N_{2}^{\mathrm{SL}}(\tau')\right]^{(l)}-\langle N_{2}^{\mathrm{SL}}(\tau')\rangle\right\}.
\end{equation}
I show in figure \ref{volprof}\subref{cohensavvolproffluc} the square root of the diagonal of the ensemble average covariance $\langle n_{2}^{\mathrm{SL}}(\tau)n_{2}^{\mathrm{SL}}(\tau')\rangle$. The square root of the diagonal $\langle n_{2}^{\mathrm{SL}}(\tau)n_{2}^{\mathrm{SL}}(\tau)\rangle$ has the interpretation as the uncertainty in the ensemble average $\langle N_{2}^{\mathrm{SL}}(\tau)\rangle$. Restricted to the central accumulation, the ensemble average covariance $\langle n_{2}^{\mathrm{SL}}(\tau)n_{2}^{\mathrm{SL}}(\tau')\rangle$ is well described by a discretization of the connected $2$-point function of metric tensor fluctuations (depending only on the global time coordinate) propagating on Euclidean de Sitter space \cite{JA&AG&JJ&RL2,JHC&KL&JMM}.

\section{Definitions}\label{definitions}

\subsection{Spacetime homogeneity}\label{spacetimehomo}

I now define two measures of the homogeneity of the quantum spacetime geometry specified by the partition function \eqref{CDTpartfunc}.

\subsubsection{Volumetric measure}


Consider a Wick-rotated causal triangulation $\mathcal{T}_{c}$ comprised of $N_{d+1}$ $(d+1)$-simplices. Select a $(d+1)$-simplex $s$ within the causal triangulation $\mathcal{T}_{c}$. Let $\mathscr{N}_{s}(r)$ be the set of $(d+1)$-simplices at a graph geodesic distance $r$ from the $(d+1)$-simplex $s$. In particular, $\mathscr{N}_{s}(0)$ is the set $\{s\}$, 
$\mathscr{N}_{s}(1)$ is the set of $d+1$ nearest neighbor $(d+1)$-simplices, $\mathscr{N}_{s}(2)$ is the set of next-nearest neighbor $(d+1)$-simplices, \emph{et cetera}. Let $N(\mathscr{N}_{s}(r))$ be the number of $(d+1)$-simplices within the set $\mathscr{N}_{s}(r)$. Now define $\mathfrak{N}_{s}(r)$ to be the total number of $(d+1)$-simplices within a graph geodesic distance $r$ from the $(d+1)$-simplex $s$ normalized by the number $N_{d+1}$ of $(d+1)$-simplices:
\begin{equation}
\mathfrak{N}_{s}(r)=\frac{1}{N_{d+1}}\sum_{j=0}^{r}N(\mathscr{N}_{s}(j)).
\end{equation}
The function $\mathfrak{N}_{s}(r)$ characterizes one particular property of the geometry of the causal triangulation $\mathcal{T}_{c}$ from the perspective of the $(d+1)$-simplex $s$: the growth in the number of neighbor $(d+1)$-simplices with the graph geodesic distance $r$.\footnote{One could define a similar quantity referring to the vertices of the causal triangulation $\mathcal{T}_{c}$; one would expect this quantity to yield identical results as the number $N_{d+1}$ of $(d+1)$-simplices increases without bound.} Let $\mathfrak{N}_{\mathcal{T}_{c}}(r)$ be the average of $\mathfrak{N}_{s}(r)$ over the causal triangulation $\mathcal{T}_{c}$:
\begin{equation}
\mathfrak{N}_{\mathcal{T}_{c}}(r)=\frac{1}{N_{d+1}}\sum_{s\in\mathcal{T}_{c}}\mathfrak{N}_{s}(r).
\end{equation}
The expectation value $\mathbb{E}_{Z_{\Sigma}[\Gamma]}[\mathfrak{N}(r)]$ of the function $\mathfrak{N}_{\mathcal{T}_{c}}(r)$ in the quantum state specified by the partition function \eqref{CDTpartfunc} is defined as
\begin{equation}
\mathbb{E}_{Z_{\Sigma}[\Gamma]}[\mathfrak{N}(r)]=\sum_{\substack{\mathcal{T}_{c} \\ \mathcal{T}_{c}\cong\Sigma\times\mathcal{I} \\ \mathcal{T}_{c}|_{\partial\mathcal{T}_{c}}=\Gamma \\ T(\mathcal{T}_{c})=\bar{T} \\ N_{d+1}(\mathcal{T}_{c})=\bar{N}_{d+1}}}\mu(\mathcal{T}_{c})\,e^{-\mathcal{S}_{\mathrm{cl}}^{(\mathrm{E})}[\mathcal{T}_{c}]/\hbar}\,\mathfrak{N}_{\mathcal{T}_{c}}(r).
\end{equation}
One defines the Hausdorff dimension $D_{\mathrm{H}}(r)$ as the $r$-dependent power with which the expectation value $\mathbb{E}_{Z_{\Sigma}[\Gamma]}[\mathfrak{N}(r)]$ scales with the graph geodesic distance $r$:
\begin{equation}
D_{\mathrm{H}}(r)=\frac{\mathrm{d}\ln{\mathbb{E}_{Z_{\Sigma}[\Gamma]}[\mathfrak{N}(r)]}}{\mathrm{d}\ln{r}}.
\end{equation}
See, for instance, \cite{JA&TB}. If the Hausdorff dimension $D_{\mathrm{H}}(r)$ equals the topological dimension $d+1$, presumably only approximately over some interval of graph geodesic distances $r$, then the quantum geometry possesses the scaling properties of a $(d+1)$-dimensional Euclidean space. 

Since the quantum geometries studied so far do not appear to approximate Euclidean space on any significant interval of scales---most certainly not on their largest scales---one does not expect the expectation value $\mathbb{E}_{Z_{\Sigma}[\Gamma]}[\mathfrak{N}(r)]$ to serve as a homogeneity measure. This issue is readily circumvented. The variance $\mathrm{var}[\mathfrak{N}_{\mathcal{T}_{c}}(r)]$ in the function $\mathfrak{N}_{\mathcal{T}_{c}}(r)$ over all $N_{d+1}$ $(d+1)$-simplices $s$ within the causal triangulation $\mathcal{T}_{c}$,
\begin{equation}
\mathrm{var}[\mathfrak{N}_{\mathcal{T}_{c}}(r)]=\frac{1}{N_{d+1}-1}\sum_{s\in\mathcal{T}_{c}}\left[\mathfrak{N}_{s}(r)-\mathfrak{N}_{\mathcal{T}_{c}}(r)\right]^{2}
\end{equation}
serves as a homogeneity measure because this variance explicitly quantities the amount by which the function $\mathfrak{N}_{s}(r)$ differs from $(d+1)$-simplex to $(d+1)$-simplex within the causal triangulation $\mathcal{T}_{c}$. In particular, if the variance $\mathrm{var}[\mathfrak{N}_{\mathcal{T}_{c}}(r)]$ vanishes for all graph geodesic distances $r$, then the causal triangulation $\mathcal{T}_{c}$ is exactly homogeneous. 
The expectation value $\mathbb{E}_{Z_{\Sigma}[\Gamma]}[\mathrm{var}[\mathfrak{N}(r)]]$ of the variance $\mathrm{var}[\mathfrak{N}_{\mathcal{T}_{c}}(r)]$ in the quantum state specified by the partition function \eqref{CDTpartfunc} is defined as
\begin{equation}
\mathbb{E}_{Z_{\Sigma}[\Gamma]}[\mathrm{var}[\mathfrak{N}(r)]]=\sum_{\substack{\mathcal{T}_{c} \\ \mathcal{T}_{c}\cong\Sigma\times\mathcal{I} \\ \mathcal{T}_{c}|_{\partial\mathcal{T}_{c}}=\Gamma \\ T(\mathcal{T}_{c})=\bar{T} \\ N_{d+1}(\mathcal{T}_{c})=\bar{N}_{d+1}}}\mu(\mathcal{T}_{c})\,e^{-\mathcal{S}_{\mathrm{cl}}^{(\mathrm{E})}[\mathcal{T}_{c}]/\hbar}\,\mathrm{var}[\mathfrak{N}_{\mathcal{T}_{c}}(r)].
\end{equation}
I thus define a scale-dependent volumetric homogeneity measure $\mathcal{H}_{V}(r)$ for the quantum spacetime geometry specified by the partition function \eqref{CDTpartfunc} as
\begin{equation}
\mathcal{H}_{V}(r)=\mathbb{E}_{Z_{\Sigma}[\Gamma]}[\mathrm{var}[\mathfrak{N}(r)]].
\end{equation}

\subsubsection{Spectral measure}


Consider the diffusion of a test random walker on a Wick-rotated causal triangulation $\mathcal{T}_{c}$. The heat equation describing this process assumes the integrated form
\begin{equation}\label{heatequation}
K(s,s',\sigma+1)=(1-\varrho)K(s,s',\sigma)+\frac{\varrho}{N(\mathscr{N}_{s}(1))}\sum_{s''\in\mathscr{N}_{s}(1)}K(s'',s',\sigma).
\end{equation}
Subject to the initial normalization condition
\begin{equation}\label{initialcondition}
K(s,s',0)=\delta_{ss'},
\end{equation}
the heat kernel $K(s,s',\sigma)$ gives the probability of diffusion from $(d+1)$-simplex $s$ to $(d+1)$-simplex $s'$ (or \emph{vice versa}) in $\sigma$ diffusion time steps. Equation \eqref{heatequation} thus dictates that the probability $K(s,s',\sigma+1)$ of diffusing from $(d+1)$-simplex $s$ to $(d+1)$-simplex $s'$ in $\sigma+1$ diffusion time steps is the weighted sum of the probability $K(s,s',\sigma)$ of having diffused from the $(d+1)$-simplex $s$ to the $(d+1)$-simplex $s'$ in $\sigma$ diffusion time steps and the probability of having diffused from $(d+1)$-simplex $s$ to a $(d+1)$-simplex $s''$ adjacent to $(d+1)$-simplex $s$ in $\sigma$ diffusion time steps. The diffusion constant $\varrho$ characterizes the dwell probability in a given diffusion time step. 

The diagonal element $K(s,s,\sigma)$ gives the probability $P_{s}(\sigma)$ for a random walker to diffuse from $(d+1)$-simplex $s$ and return to $(d+1)$-simplex $s$ in $\sigma$ diffusion time steps. The probability $P_{s}(\sigma)$ characterizes one particular property of the geometry of the causal triangulation $\mathcal{T}_{c}$ from the perspective of the $(d+1)$-simplex $s$: the likelihood that a random walker returns in $\sigma$ diffusion time steps. Let $P_{\mathcal{T}_{c}}(\sigma)$ be the average of $P_{s}(\sigma)$ over the causal triangulation $\mathcal{T}_{c}$:
\begin{equation}
P_{\mathcal{T}_{c}}(\sigma)=\frac{1}{N_{d+1}}\sum_{s\in\mathcal{T}_{c}}P_{s}(\sigma).
\end{equation}
The expectation value $\mathbb{E}_{Z_{\Sigma}[\Gamma]}[P(\sigma)]$ of the probability $P_{\mathcal{T}_{c}}(\sigma)$ in the quantum state specified by the partition function \eqref{CDTpartfunc} is defined as
\begin{equation}
\mathbb{E}_{Z_{\Sigma}[\Gamma]}[P(\sigma)]=\sum_{\substack{\mathcal{T}_{c} \\ \mathcal{T}_{c}\cong\Sigma\times\mathcal{I} \\ \mathcal{T}_{c}|_{\partial\mathcal{T}_{c}}=\Gamma \\ T(\mathcal{T}_{c})=\bar{T} \\ N_{d+1}(\mathcal{T}_{c})=\bar{N}_{d+1}}}\mu(\mathcal{T}_{c})\,e^{-\mathcal{S}_{\mathrm{cl}}^{(\mathrm{E})}[\mathcal{T}_{c}]/\hbar}\,P_{\mathcal{T}_{c}}(\sigma).
\end{equation}
One defines the spectral dimension $D_{S}(\sigma)$ as the (negative of twice the) $\sigma$-dependent power with which the expectation value $\mathbb{E}_{Z_{\Sigma}[\Gamma]}[P(\sigma)]$ scales with the diffusion time $\sigma$:
\begin{equation}
D_{S}(\sigma)=-2\frac{\mathrm{d}\ln{\mathbb{E}_{Z_{\Sigma}[\Gamma]}[P(\sigma)]}}{\mathrm{d}\ln{\sigma}}.
\end{equation}
If the spectral dimension $D_{S}(\sigma)$ equals the topological dimension $d+1$, again presumably only approximately over some interval of scales, then the quantum geometry possesses certain propagation properties of a $(d+1)$-dimensional Euclidean space. 

Since the quantum geometries studied so far do only appear to approximate a Euclidean space on any significant interval of scales---most certainly not on their largest scales---one does not expect the expectation value $\mathbb{E}[P(\sigma)]$ to serve as a homogeneity measure. This issue is again readily circumvented. The variance $\mathrm{var}[P_{\mathcal{T}_{c}}(\sigma)]$ in the probability $P_{\mathcal{T}_{c}}(\sigma)$ over all $N_{d+1}$ $(d+1)$-simplices within the causal triangulation $\mathcal{T}_{c}$,
\begin{equation}
\mathrm{var}[P_{\mathcal{T}_{c}}(\sigma)]=\frac{1}{N_{d+1}-1}\sum_{s\in\mathcal{T}_{c}}\left[P_{s}(\sigma)-P_{\mathcal{T}_{c}}(\sigma)\right]^{2},
\end{equation}
serves as a homogeneity measure because this variance explicitly quantities the amount by which the probability $P_{s}(\sigma)$ differs from $(d+1)$-simplex to $(d+1)$-simplex within the causal triangulation $\mathcal{T}_{c}$. In particular, if the variance $\mathrm{var}[P_{\mathcal{T}_{c}}(\sigma)]$ vanishes for all diffusion times $\sigma$, then the causal triangulation $\mathcal{T}_{c}$ is exactly homogeneous. The expectation value $\mathbb{E}_{Z_{\Sigma}[\Gamma]}[\mathrm{var}[P(\sigma)]]$ of the variance $\mathrm{var}[P_{\mathcal{T}_{c}}(\sigma)]$ in the quantum state specified by the partition function \eqref{CDTpartfunc} is defined as
\begin{equation}
\mathbb{E}_{Z_{\Sigma}[\Gamma]}[\mathrm{var}[P(\sigma)]]=\sum_{\substack{\mathcal{T}_{c} \\ \mathcal{T}_{c}\cong\Sigma\times\mathcal{I} \\ \mathcal{T}_{c}|_{\partial\mathcal{T}_{c}}=\Gamma \\ T(\mathcal{T}_{c})=\bar{T} \\ N_{d+1}(\mathcal{T}_{c})=\bar{N}_{d+1}}}\mu(\mathcal{T}_{c})\,e^{-\mathcal{S}_{\mathrm{cl}}^{(\mathrm{E})}[\mathcal{T}_{c}]/\hbar}\,\mathrm{var}[P_{\mathcal{T}_{c}}(\sigma)].
\end{equation}
I thus define a scale-dependent spectral homogeneity measure $H_{S}(\sigma)$ for the quantum spacetime geometry specified by the partition function \eqref{CDTpartfunc} as
\begin{equation}
H_{S}(\sigma)=\mathbb{E}_{Z_{\Sigma}[\Gamma]}[\mathrm{var}[P(\sigma)]].
\end{equation}

\subsection{Temporal evolution of spatial homogeneity}\label{spatialhomo}

I next define two measures of the homogeneity of the quantum spatial geometry specified by the partition function \eqref{CDTpartfunc}. By considering the temporal evolution of the quantum spatial geometry, I then trace the temporal evolution of the homogeneity of the quantum spatial geometry. To formulate these definitions, I require a means firstly to identify spatial geometries within a causal triangulation and subsequently to identify spatial geometries across causal triangulations. 
I now explain how to accomplish these ends.

To identify spatial geometries, I employ the distinguished foliation of a causal triangulation. This foliation provides a primitive notion of spatial geometry: at any given value of the discrete time coordinate $\tau$, there is a leaf of this foliation constructed entirely from regular spacelike $d$-simplices. 
I assume that each such leaf constitutes a spatial geometry not only in this classical sense, but also at the quantum level. This assumption is consistent with the interpretation of the quantum geometry of phase C on large scales as that of Euclidean de Sitter space. To identify leaves of the distinguished foliation across causal triangulations, I employ the 
technique of aligning the zero of the discrete time coordinate with the center of discrete spatial $2$-volume explained in subsection \ref{phenomenology}.

Equipped with the means to identify spatial geometries across all of the causal triangulations in an ensemble, I now define the two homogeneity measures of the quantum spatial geometry in complete analogy to those of subsection \ref{spacetimehomo}.




\subsubsection{Volumetric measure}

Consider a distinguished spacelike $d$-surface $\mathsf{T}_{\tau}$ of a causal triangulation $\mathcal{T}_{c}$ labeled by the discrete time coordinate $\tau$ and comprised of $N_{d}$ spacelike $d$-simplices $\mathsf{s}$. Select a $d$-simplex $\mathsf{s}$ within this spacelike $d$-surface. Let $\mathcal{N}_{\mathsf{s}}(r)$ be the set of $d$-simplices at a graph geodesic distance $r$ from the $d$-simplex $\mathsf{s}$. In particular, $\mathcal{N}_{\mathsf{s}}(0)$ is the set $\{\mathsf{s}\}$, $\mathcal{N}_{\mathsf{s}}(1)$ is the set of $d$ nearest neighbor $d$-simplices, $\mathcal{N}_{\mathsf{s}}(2)$ is the set of next-nearest neighbor $d$-simplices, \emph{et cetera}. Let $N(\mathcal{N}_{\mathsf{s}}(r))$ be the number of $d$-simplices within the set $\mathcal{N}_{\mathsf{s}}(r)$. Now define $\mathsf{N}_{\mathsf{s}}(r)$ to be the total number of $d$-simplices within a graph geodesic distance $r$ from the $d$-simplex $\mathsf{s}$ normalized by the number $N_{d}$ of $d$-simplices:
\begin{equation}
\mathsf{N}_{\mathsf{s}}(r)=\frac{1}{N_{d}}\sum_{j=0}^{r}N(\mathcal{N}_{\mathsf{s}}(j)).
\end{equation}
The function $\mathsf{N}_{\mathsf{s}}(r)$ characterizes one particular property of the geometry of the spacelike $d$-surface $\mathsf{T}_{\tau}$ from the perspective of the $d$-simplex $\mathsf{s}$: the growth in the number of neighbor $d$-simplices with the graph geodesic distance $r$. Let $\mathsf{N}_{\mathsf{T}_{\tau}}(r)$ be the average of $\mathsf{N}_{\mathsf{s}}(r)$ over the spacelike $d$-surface $\mathsf{T}_{\tau}$:
\begin{equation}
\mathsf{N}_{\mathsf{T}_{\tau}}(r)=\frac{1}{N_{d}}\sum_{\mathsf{s}\in\mathsf{T}_{\tau}}\mathsf{N}_{\mathsf{s}}(r).
\end{equation}
The expectation value $\mathbb{E}_{Z_{\Sigma}[\Gamma]}[\mathsf{N}_{\mathsf{T}_{\tau}}(r)]$ of the function $\mathsf{N}_{\mathsf{T}_{\tau}}(r)$ in the quantum state specified by the partition function \eqref{CDTpartfunc} is defined as
\begin{equation}\label{xvalueTESHvol}
\mathbb{E}_{Z_{\Sigma}[\Gamma]}[\mathsf{N}_{\mathsf{T}_{\tau}}(r)]=\sum_{\substack{\mathcal{T}_{c} \\ \mathcal{T}_{c}\cong\Sigma\times\mathcal{I} \\ \mathcal{T}_{c}|_{\partial\mathcal{T}_{c}}=\Gamma \\ T(\mathcal{T}_{c})=\bar{T} \\ N_{d+1}(\mathcal{T}_{c})=\bar{N}_{d+1}}}\mu(\mathcal{T}_{c})\,e^{-\mathcal{S}_{\mathrm{cl}}^{(\mathrm{E})}[\mathcal{T}_{c}]/\hbar}\,\mathsf{N}_{\mathsf{T}_{\tau}}^{(\mathcal{T}_{c})}(r).
\end{equation}
One defines the Hausdorff dimension $d_{\mathrm{H}}(r)$ as the $r$-dependent power with which the expectation value $\mathbb{E}_{Z_{\Sigma}[\Gamma]}[\mathsf{N}_{\mathsf{T}_{\tau}}(r)]$ scales with the graph geodesic distance $r$:
\begin{equation}\label{HdimTESHvol}
d_{\mathrm{H}}(r)=\frac{\mathrm{d}\ln{\mathbb{E}_{Z_{\Sigma}[\Gamma]}[\mathsf{N}_{\mathsf{T}_{\tau}}(r)]}}{\mathrm{d}\ln{r}}.
\end{equation}
If the Hausdorff dimension $d_{\mathrm{H}}(r)$ equals the topological dimension $d$, presumably only approximately over some interval of graph geodesic distances $r$, then the quantum geometry of the spacelike $d$-surface $\mathsf{T}_{\tau}$ possesses the scaling properties of a $d$-dimensional Euclidean space.

The variance $\mathrm{var}[\mathsf{N}_{\mathsf{T}_{\tau}}(r)]$ in the function $\mathsf{N}_{\mathsf{T}_{\tau}}(r)$ over all $N_{d}$ $d$-simplices $\mathsf{s}$ within the spacelike $d$-surface $\mathsf{T}_{\tau}$,
\begin{equation}
\mathrm{var}[\mathsf{N}_{\mathsf{T}_{\tau}}(r)]=\frac{1}{N_{d}-1}\sum_{\mathsf{s}\in\mathsf{T}_{\tau}}\left[\mathsf{N}_{\mathsf{s}}(r)-\mathsf{N}_{\mathsf{T}_{\tau}}(r)\right]^{2},
\end{equation}
serves as a homogeneity measure because this variance explicitly quantifies the amount by which the function $\mathsf{N}_{\mathsf{s}}(r)$ differs from $d$-simplex to $d$-simplex within the spacelike $d$-surface $\mathsf{T}_{\tau}$. In particular, if the variance $\mathrm{var}[\mathsf{N}_{\mathsf{T}_{\tau}}(r)]$ vanishes for all graph geodesic distances $r$, then the spacelike $d$-surface $\mathsf{T}_{\tau}$ is exactly homogeneous. The expectation value $\mathbb{E}_{Z_{\Sigma}[\Gamma]}[\mathrm{var}[\mathsf{N}_{\mathsf{T}_{\tau}}(r)]]$ of the variance $\mathrm{var}[\mathsf{N}_{\mathsf{T}_{\tau}}(r)]$ in the quantum state specified by the partition function \eqref{CDTpartfunc} is defined as
\begin{equation}
\mathbb{E}_{Z_{\Sigma}[\Gamma]}[\mathrm{var}[\mathsf{N}_{\mathsf{T}_{\tau}}(r)]]=\sum_{\substack{\mathcal{T}_{c} \\ \mathcal{T}_{c}\cong\Sigma\times\mathcal{I} \\ \mathcal{T}_{c}|_{\partial\mathcal{T}_{c}}=\Gamma \\ T(\mathcal{T}_{c})=\bar{T} \\ N_{d+1}(\mathcal{T}_{c})=\bar{N}_{d+1}}}\mu(\mathcal{T}_{c})\,e^{-\mathcal{S}_{\mathrm{cl}}^{(\mathrm{E})}[\mathcal{T}_{c}]/\hbar}\,\mathrm{var}[\mathsf{N}_{\mathsf{T}_{\tau}}^{(\mathcal{T}_{c})}(r). 
\end{equation}
I thus define a scale-dependent volumetric homogeneity measure $H_{V}(r)$ for the quantum geometry of the spacelike $d$-surface $\mathsf{T}_{\tau}$ specified by the partition function \eqref{CDTpartfunc} as
\begin{equation}
H_{V}(r)=\mathbb{E}_{Z_{\Sigma}[\Gamma]}[\mathrm{var}[\mathsf{N}_{\mathsf{T}_{\tau}}(r)]].
\end{equation}
To study the temporal evolution of the homogeneity measure $H_{V}(r)$, I simply consider the succession of spacelike $d$-surfaces $\mathsf{T}_{\tau}$ for successive values of the discrete time coordinate $\tau$. 

\subsubsection{Spectral measure}

Consider the diffusion of a test random walker on a distinguished spacelike $d$-surface $\mathsf{T}_{\tau}$ of a causal triangulation $\mathcal{T}_{c}$. The heat equation describing this process assumes the integrated form
\begin{equation}\label{spatialheatequation}
\mathcal{K}(\mathsf{s},\mathsf{s}',\sigma+1)=(1-\rho)\mathcal{K}(\mathsf{s},\mathsf{s}',\sigma)+\frac{\rho}{N(\mathscr{N}_{\mathsf{s}}(1))}\sum_{\mathsf{s}''\in\mathscr{N}_{\mathsf{s}}(1)}\mathcal{K}(\mathsf{s}'',\mathsf{s}',\sigma).
\end{equation}
Subject to the initial normalization condition 
\begin{equation}
\mathcal{K}(\mathsf{s},\mathsf{s}',0)=\delta_{\mathsf{s}\mathsf{s}'},
\end{equation}
the heat kernel $\mathcal{K}(\mathsf{s},\mathsf{s}',\sigma)$ gives the probability of diffusion from $d$-simplex $\mathsf{s}$ to $d$-simplex $\mathsf{s}'$ (or \emph{vice versa}) in $\sigma$ diffusion time steps. Equation \eqref{spatialheatequation} thus dictates that the probability $\mathcal{K}(\mathsf{s},\mathsf{s}',\sigma+1)$ of diffusing from $d$-simplex $\mathsf{s}$ to $d$-simplex $\mathsf{s}'$ in $\sigma+1$ diffusion time steps is the weighted sum of the probability $\mathcal{K}(\mathsf{s},\mathsf{s}',\sigma)$ of having diffused from $d$-simplex $\mathsf{s}$ to $d$-simplex $\mathsf{s}'$ in $\sigma$ diffusion time steps and the probability of having diffused from $d$-simplex $\mathsf{s}$ to a $d$-simplex $\mathsf{s}''$ adjacent to $d$-simplex $\mathsf{s}$ in $\sigma$ diffusion time steps. The diffusion constant $\varrho$ characterizes the dwell probability in a given diffusion time step. 

The diagonal element $\mathcal{K}(\mathsf{s},\mathsf{s},\sigma)$ gives the probability $\mathcal{P}_{s}(\sigma)$ for a random walker to diffuse from $d$-simplex $\mathsf{s}$ and return to $d$-simplex $\mathsf{s}$ in $\sigma$ diffusion time steps. The probability $\mathcal{P}_{\mathsf{s}}(\sigma)$ characterizes one particular property of the geometry of the spacelike $d$-surface $\mathsf{T}_{\tau}$ from the perspective of the $d$-simplex $\mathsf{s}$: the likelihood that a random walker returns in $\sigma$ diffusion time steps. Let $\mathcal{P}_{\mathsf{T}_{\tau}}(\sigma)$ be the average of $\mathcal{P}_{\mathsf{s}}(\sigma)$ over the spacelike $d$-surface $\mathsf{T}_{\tau}$:
\begin{equation}
\mathcal{P}_{\mathsf{T}_{\tau}}(\sigma)=\frac{1}{N_{d}}\sum_{\mathsf{s}\in\mathsf{T}_{\tau}}\mathcal{P}_{\mathsf{s}}(\sigma).
\end{equation}
The expectation value $\mathbb{E}_{Z_{\Sigma}[\Gamma]}[\mathcal{P}_{\mathsf{T}_{\tau}}(\sigma)]$ of the probability $\mathcal{P}_{\mathsf{T}_{\tau}}(\sigma)$ in the quantum state specified by the partition function \eqref{CDTpartfunc} is defined as
\begin{equation}
\mathbb{E}_{Z_{\Sigma}[\Gamma]}[\mathcal{P}_{\mathsf{T}_{\tau}}(\sigma)]=\sum_{\substack{\mathcal{T}_{c} \\ \mathcal{T}_{c}\cong\Sigma\times\mathcal{I} \\ \mathcal{T}_{c}|_{\partial\mathcal{T}_{c}}=\Gamma \\ T(\mathcal{T}_{c})=\bar{T} \\ N_{d+1}(\mathcal{T}_{c})=\bar{N}_{d+1}}}\mu(\mathcal{T}_{c})\,e^{-\mathcal{S}_{\mathrm{cl}}^{(\mathrm{E})}[\mathcal{T}_{c}]/\hbar}\,\mathcal{P}_{\mathsf{T}_{\tau}}^{(\mathcal{T}_{c})}(\sigma).
\end{equation}
One defines the spectral dimension $d_{S}(\sigma)$ as the (negative of twice the) $\sigma$-dependent power with which the expectation value $\mathbb{E}_{Z_{\Sigma}[\Gamma]}[\mathcal{P}_{\mathsf{T}_{\tau}}(\sigma)]$ scales with the diffusion time $\sigma$:
\begin{equation}
d_{S}(\sigma)=-2\frac{\mathrm{d}\ln{\mathbb{E}_{Z_{\Sigma}[\Gamma]}[\mathcal{P}_{\mathsf{T}_{\tau}}(\sigma)]}}{\mathrm{d}\ln{\sigma}}.
\end{equation}
If the spectral dimension $d_{S}(\sigma)$ equals the topological dimension $d$, again presumably only approximately over some interval of diffusion times, then the quantum geometry possesses certain propagation properties of a $d$-dimensional Euclidean space. 

The variance $\mathrm{var}[\mathcal{P}_{\mathsf{T}_{\tau}}(\sigma)]$ in the probability $\mathcal{P}_{\mathsf{T}_{\tau}}(\sigma)$ over all $N_{d}$ $d$-simplices within the spacelike $d$-surface $\mathsf{T}_{\tau}$,
\begin{equation}
\mathrm{var}[\mathcal{P}_{\mathsf{T}_{\tau}}(\sigma)]=\frac{1}{N_{d}-1}\sum_{\mathsf{s}\in\mathsf{T}_{\tau}}\left[\mathcal{P}_{\mathsf{s}}(\sigma)-\mathcal{P}_{\mathsf{T}_{\tau}}(\sigma)\right]^{2},
\end{equation}
serves as a homogeneity measure because this variance explicitly quantities the amount by which the probability $\mathcal{P}_{\mathsf{s}}(\sigma)$ differs from $d$-simplex to $d$-simplex within the spacelike $d$-surface $\mathsf{T}_{\tau}$. In particular, if the variance $\mathrm{var}[\mathcal{P}_{\mathsf{T}_{\tau}}(\sigma)]$ vanishes for all diffusion times $\sigma$, then the spacelike $d$-surface $\mathsf{T}_{\tau}$ is exactly homogeneous. The expectation value $\mathbb{E}_{Z_{\Sigma}[\Gamma]}[\mathrm{var}[\mathcal{P}_{\mathsf{T}_{\tau}}(\sigma)]]$ of the variance $\mathrm{var}[\mathcal{P}_{\mathsf{T}_{\tau}}(\sigma)]$ in the quantum state specified by the partition function \eqref{CDTpartfunc} is defined as
\begin{equation}
\mathbb{E}_{Z_{\Sigma}[\Gamma]}[\mathrm{var}[\mathcal{P}_{\mathsf{T}_{\tau}}(\sigma)]]=\sum_{\substack{\mathcal{T}_{c} \\ \mathcal{T}_{c}\cong\Sigma\times\mathcal{I} \\ \mathcal{T}_{c}|_{\partial\mathcal{T}_{c}}=\Gamma \\ T(\mathcal{T}_{c})=\bar{T} \\ N_{d+1}(\mathcal{T}_{c})=\bar{N}_{d+1}}}\mu(\mathcal{T}_{c})\,e^{-\mathcal{S}_{\mathrm{cl}}^{(\mathrm{E})}[\mathcal{T}_{c}]/\hbar}\,\mathrm{var}[\mathcal{P}_{\mathsf{T}_{\tau}}^{(\mathcal{T}_{c})}(\sigma)].
\end{equation}
I thus define a scale-dependent spectral homogeneity measure $H_{S}(\sigma)$ for the quantum geometry of the spacelike $d$-surface $\mathsf{T}_{\tau}$ specified by the partition function \eqref{CDTpartfunc} as
\begin{equation}
H_{S}(\sigma)=\mathbb{E}_{Z_{\Sigma}[\Gamma]}[\mathrm{var}[\mathcal{P}_{\mathsf{T}_{\tau}}(\sigma)]].
\end{equation}
To study the temporal evolution of the homogeneity measure $H_{S}(\sigma)$, I simply consider the succession of spacelike $d$-surfaces $\mathsf{T}_{\tau}$ for successive values of the discrete time coordinate $\tau$.





\section{Results}\label{results}

\subsection{Spacetime homogeneity}\label{spacetimehomoresults}

I have measured the homogeneity measures $\mathcal{H}_{V}(r)$ and $\mathcal{H}_{S}(\sigma)$ for three ensembles of causal triangulations characterized by the same bare couplings within phase C for increasing values of the number $N_{3}$ of $3$-simplices. When estimating the homogeneity measures $\mathcal{H}_{V}(r)$ and $\mathcal{H}_{S}(\sigma)$ as explained in appendix \ref{appendix}, I restrict consideration to those leaves of the distinguished foliation that fall within the central accumulation. I first display the results for each ensemble, and I then perform a finite size scaling analysis of all three ensembles. Through a finite size scaling analysis, one attempts to extrapolate the limit of arbitrarily large $N_{3}$, thereby removing the influence of finite $N_{3}$ from the measurements of a discrete observable. 
This extrapolation does not necessarily provide information about the continuum limit of a discrete observable as the continuum limit additionally involves the limit of vanishing lattice spacing. 

\subsubsection{Volumetric measure}\label{spacetimehomoresultsvol}

\begin{figure}[h!]
\centering
\subfigure[ ]{
\includegraphics[scale=0.875]{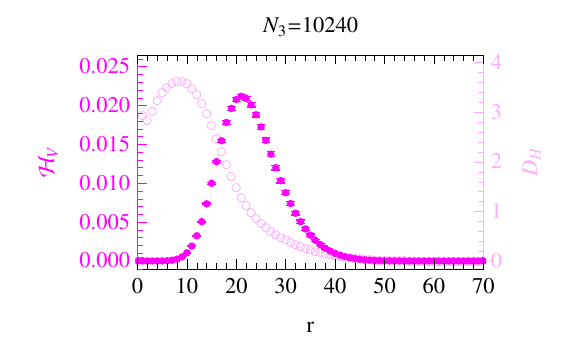}
\label{volhomogeneity10}
}
\subfigure[ ]{
\includegraphics[scale=0.875]{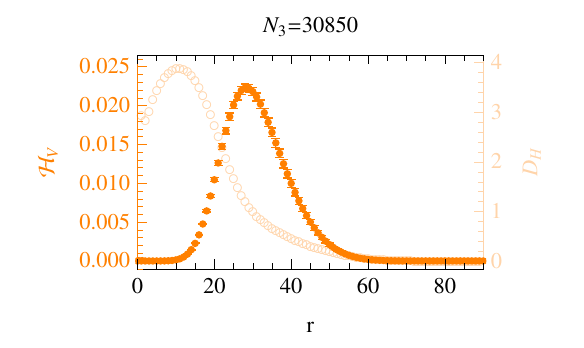}
\label{volhomogeneity30}
}
\subfigure[ ]{
\includegraphics[scale=0.875]{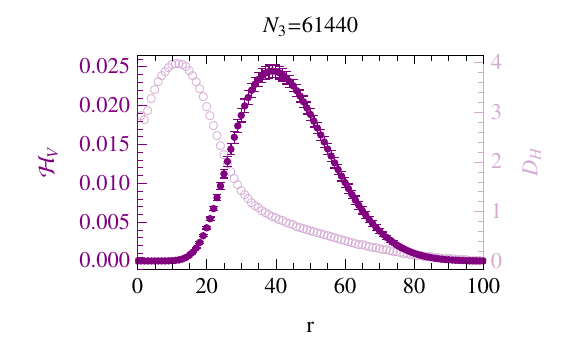}
\label{volhomogeneity60}
}
\caption[Optional caption for list of figures]{Volumetric homogeneity measure $\mathcal{H}_{V}$ (dark, left axis) and Hausdorff dimension $D_{\mathrm{H}}$ (light, right axis) of the quantum spacetime geometry as a function of the graph geodesic distance $r$ for ensembles of causal triangulations characterized by $T=64$ and $k_{0}=1.0$. \subref{volhomogeneity10} $N_{3}=10850$ (magenta). \subref{volhomogeneity30} $N_{3}=30850$ (orange). \subref{volhomogeneity60} $N_{3}=61440$ (purple).}
\label{volhomogeneity}
\end{figure}

\begin{figure}[h!]
\centering
\subfigure[ ]{
\includegraphics[scale=1]{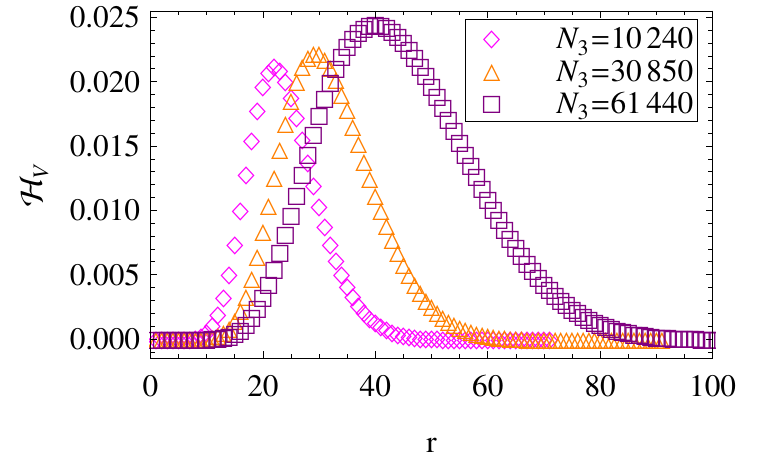}
\label{allvol}
}
\subfigure[ ]{
\includegraphics[scale=1]{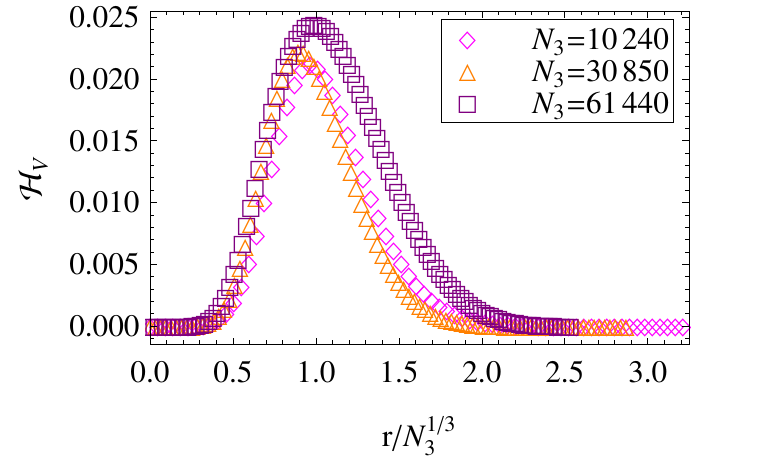}
\label{FSSvol}
}
\caption[Optional caption for list of figures]{\subref{allvol} Volumetric homogeneity measure $\mathcal{H}_{V}$ of the quantum spacetime geometry as a function of the graph geodesic distance $r$ for three ensembles of causal triangulations characterized by $T=64$ and $k_{0}=1.0$. \subref{FSS} Finite size scaling analysis of the volumetric homogeneity measure $\mathcal{H}_{V}$ of the quantum spacetime geometry as a function of the graph geodesic distance $r$ for three ensembles of causal triangulations characterized by $T=64$ and $k_{0}=1.0$.}
\label{volhomogeneityFSS}
\end{figure}

In figure \ref{volhomogeneity} I display measurements of the volumetric homogeneity measure $\mathcal{H}_{V}(r)$ and, as a reference, the Hausdorff dimension $D_{\mathrm{H}}(r)$ for three ensembles of causal triangulations at fixed number $T$ 
and fixed coupling $k_{0}$.\footnote{The Hausdorff dimension $D_{\mathrm{H}}(r)$ has not previously been studied for ensembles of causal triangulations within phase C. Its measured values are clearly of the correct order of magnitude, but I defer further consideration to future work.}
For ease of comparison, in figure \ref{volhomogeneityFSS}\subref{allvol} I display all three of these measurements together. 

The shape of the homogeneity measure $\mathcal{H}_{V}(r)$ is readily explained. The homogeneity measure $\mathcal{H}_{V}(r)$ vanishes exactly at $r=0$ by definition and at $r=1$ because all $3$-simplices have four nearest neighbor $3$-simplices.\footnote{Technically, those $(3,1)$ $3$-simplices with three vertices on the initial leaf of the distinguished foliation within the central accumulation and those $(1,3)$ $3$-simplices with three vertices on the final leaf of the distinguished foliation within the central accumulation have only three nearest neighbor $3$-simplices. Since these $3$-simplices represent a negligible fraction of all $N_{3}$ $3$-simplices, the homogeneity measure $\mathcal{H}_{V}(r)$ differs only negligibly from zero at $r=1$.} As the graph geodesic distance $r$ subsequently increases, the possible numbers of neighbor $3$-simplices within the graph geodesic distance $r$ also increases initially, so the homogeneity measure $\mathcal{H}_{V}(r)$ increases initially as well. As the graph geodesic distance $r$ further increases, however, the possible numbers of neighbor $3$-simplices within the graph geodesic distance $r$ decreases, so the homogeneity measure $\mathcal{H}_{V}(r)$ decreases as well. These decreases stem from the fact that I consider closed causal triangulations. 
The homogeneity measure $\mathcal{H}_{V}(r)$ again vanishes exactly for sufficiently large graph geodesic distances $r$ because all $N_{3}$ $3$-simplices eventually fall within some finite graph geodesic distance $r_{\mathrm{comp}}$. 

One could also have inferred the shape of the homogeneity measure $\mathcal{H}_{V}(r)$ from continuity given its values at $r=0$ and $r=r_{\mathrm{comp}}$. In this sense the homogeneity measure $\mathcal{H}_{V}(r)$ does not capture particularly well the inhomogeneity of the quantum spacetime geometry for small and large graph geodesic distances. Of course, for sufficiently small graph geodesic distances, one does not expect the homogeneity measure $\mathcal{H}_{V}(r)$ to provide reliable results since it largely probes the discreteness of the quantum spacetime geometry on these scales. 
For intermediate graph geodesic distances, on which the homogeneity measure does capture more reliably the inhomogeneity of the quantum spacetime geometry, there are two pieces of information not dictated by continuity alone: the graph geodesic distance $r_{\mathrm{max}}$ at which the homogeneity measure $\mathcal{H}_{V}(r)$ attains its maximum value and the value $\mathcal{H}_{V}(r_{\mathrm{max}})$ of the homogeneity measure at the graph geodesic distance $r_{\mathrm{max}}$.  
These values contain the essential information about the inhomogeneity of the quantum spacetime geometry according to the homogeneity measure $\mathcal{H}_{V}(r)$. 

In figure \ref{volhomogeneityFSS}\subref{FSSvol} I display the results of a finite size scaling analysis of the volumetric homogeneity measure $\mathcal{H}_{V}(r)$ for the same three ensembles of causal triangulations.
The homogeneity measure $\mathcal{H}_{V}(r)$ is dimensionless, so one does not expect it to finite size scale. The graph geodesic distance is associated with the dimensions of length, so one expects it to finite size scale as $r/N_{3}^{1/3}$. The approximate equivalence of the finite size scaled graph geodesic distance $r_{\mathrm{max}}/N_{3}^{1/3}$, defined below, for the three ensembles, quantified in figure \ref{volhomogeneityFSS2}\subref{volhomogeneitymaxscalevsN3FSS} below, supports this expectation. 

I wish to quantify the typical scale, as measured by the graph geodesic distance $r$, on which the quantum spacetime geometry exhibits inhomogeneity and the typical magnitude, as measured by the homogeneity measure $\mathcal{H}_{V}(r)$, of inhomogeneity of the quantum spacetime geometry on this typical scale. Given the shape of the homogeneity measure $\mathcal{H}_{V}(r)$,  
I take the graph geodesic distance $r_{\mathrm{max}}$ at the maximum of the homogeneity measure $\mathcal{H}_{V}(r)$ as the typical scale of inhomogeneity and 
the homogeneity measure $\mathcal{H}_{V}(r_{\mathrm{max}})$ at the graph geodesic distance $r_{\mathrm{max}}$ as the typical magnitude of inhomogeneity. I now consider three particular aspects of the finite size scaling analysis. In figure \ref{volhomogeneityFSS2}\subref{volhomogeneitymaxscalevsN3} I display the graph geodesic distance $r_{\mathrm{max}}$ as a function of the number $N_{3}$ of $3$-simplices, which evidently attests to a positive linear relation. 
\begin{figure}[h!]
\centering
\subfigure[ ]{
\includegraphics[scale=0.625]{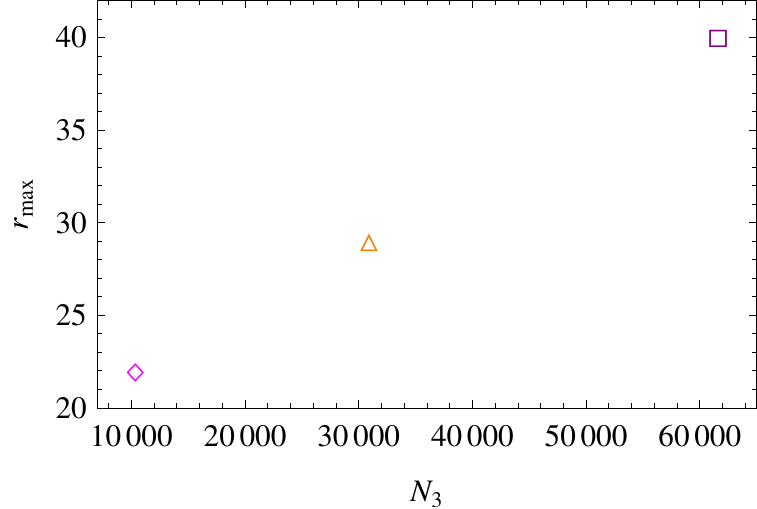}
\label{volhomogeneitymaxscalevsN3}
}
\subfigure[ ]{
\includegraphics[scale=0.65]{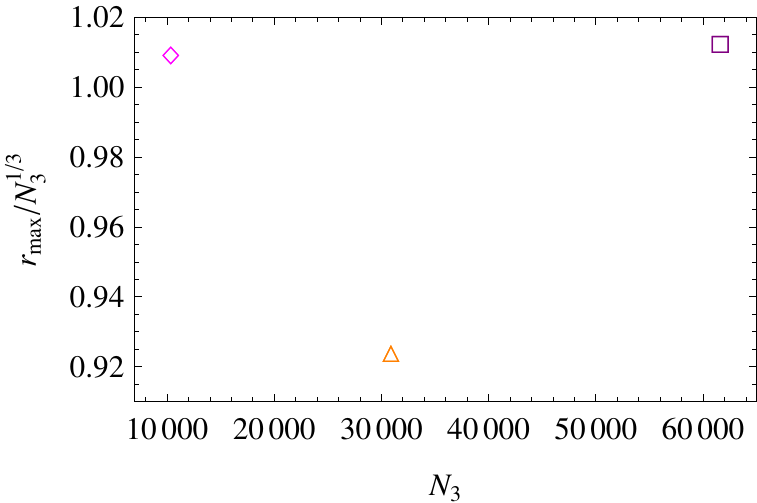}
\label{volhomogeneitymaxscalevsN3FSS}
}
\subfigure[ ]{
\includegraphics[scale=0.65]{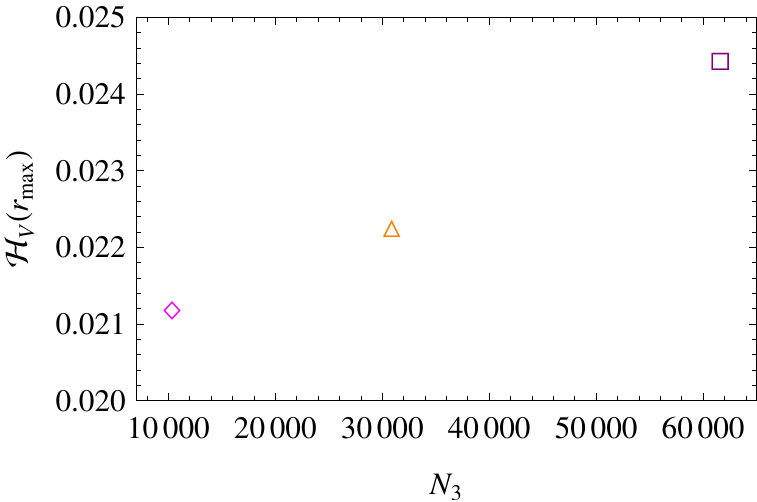}
\label{volhomogeneitymaxmagvsN3}
}
\caption{\subref{volhomogeneitymaxscalevsN3} The graph geodesic distance $r_{\mathrm{max}}$ as a function of the number $N_{3}$ of $3$-simplices for three ensembles of causal triangulations characterized by $T=64$ and $k_{0}=1.0$. \subref{volhomogeneitymaxscalevsN3FSS} The finite size scaled graph geodesic distance $r_{\mathrm{max}}/N_{3}^{1/3}$ as a function of the number $N_{3}$ of $3$-simplices for three ensembles of causal triangulations characterized by $T=64$ and $k_{0}=1.0$. \subref{volhomogeneitymaxmagvsN3} The volumetric homogeneity measure $\mathcal{H}_{V}(r_{\mathrm{max}})$ of the quantum spacetime geometry as a function of the number $N_{3}$ of $3$-simplices for three ensembles of causal triangulations characterized by $T=64$ and $k_{0}=1.0$.}
\label{volhomogeneityFSS2}
\end{figure}
In figure \ref{volhomogeneityFSS2}\subref{volhomogeneitymaxscalevsN3FSS} I display the finite size scaled graph geodesic distance $r_{\mathrm{max}}/N_{3}^{1/3}$ as a function of the number $N_{3}$ of $3$-simplices, which suggests an approximately constant relation. 
In figure \ref{volhomogeneityFSS2}\subref{volhomogeneitymaxmagvsN3} I display the homogeneity measure $\mathcal{H}_{V}(r_{\mathrm{max}})$ as a function of the number $N_{3}$ of $3$-simplices, which also evidently attests to a positive linear relation. 

If the quantum spacetime geometry is exactly homogeneous on all scales in the limit of arbitrarily large $N_{3}$, then this property would manifest itself as a decreasing value of the homogeneity measure $\mathcal{H}_{V}(r_{\mathrm{max}})$ with increasing number $N_{3}$ of $3$-simplices. The finite size scaling analysis indicates that the opposite occurs: the homogeneity measure $\mathcal{H}_{V}(r_{\mathrm{max}})$ increases with the number $N_{3}$ of $3$-simplices. 
This finding suggests that the homogeneity measure $\mathcal{H}_{V}(r)$ is not well-defined in the limit of arbitrarily large $N_{3}$. Whether or not the homogeneity measure is well-defined in the continuum limit is another question, although this finding hints that it may not be. 

\subsubsection{Spectral measure}\label{spacetimehomoresultsspec}

\begin{figure}[h!]
\centering
\subfigure[ ]{
\includegraphics[scale=0.875]{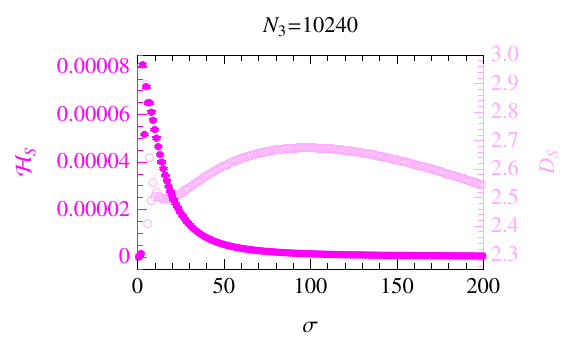}
\label{spechomogeneity10}
}
\subfigure[ ]{
\includegraphics[scale=0.875]{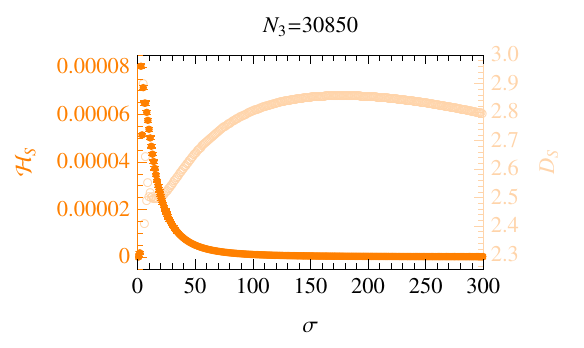}
\label{spechomogeneity30}
}
\subfigure[ ]{
\includegraphics[scale=0.875]{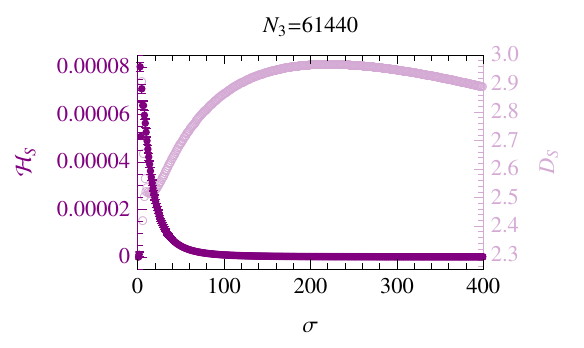}
\label{spechomogeneity60}
}
\caption[Optional caption for list of figures]{Spectral homogeneity measure $\mathcal{H}_{S}$ (black, left axis) and spectral dimension $D_{\mathrm{S}}$ (grey, right axis) of the quantum spacetime geometry as a function of the diffusion time $\sigma$ for an ensemble of causal triangulations characterized by $T=64$ and $k_{0}=1.0$. \subref{spechomogeneity10} $N_{3}=10850$ \subref{spechomogeneity30} $N_{3}=30850$ \subref{spechomogeneity60} $N_{3}=61440$}
\label{spechomogeneity}
\end{figure}

\begin{figure}[h!]
\centering
\subfigure[ ]{
\includegraphics[scale=1]{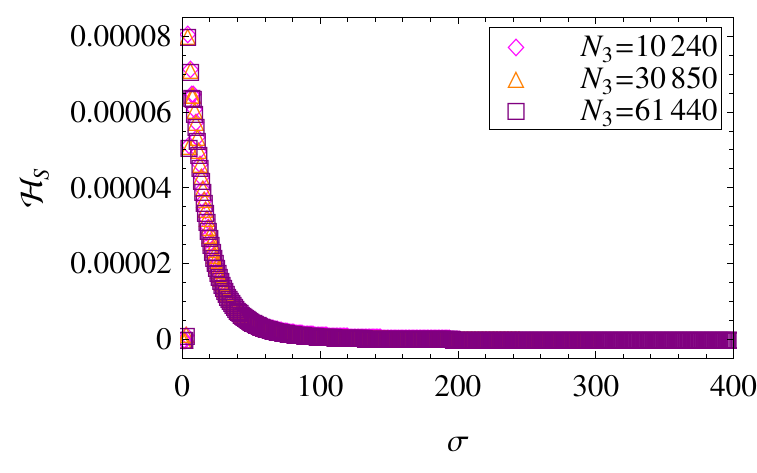}
\label{allspec}
}
\subfigure[ ]{
\includegraphics[scale=1]{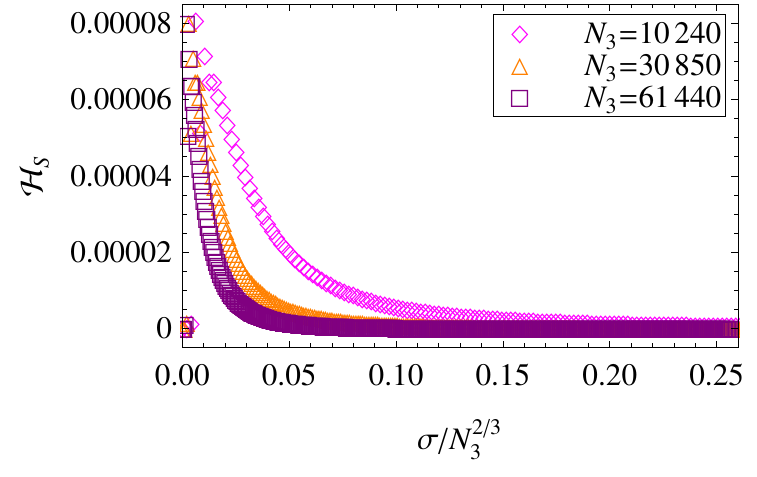}
\label{FSSspec}
}
\caption[Optional caption for list of figures]{\subref{allspec} Spectral homogeneity measure $\mathcal{H}_{S}$ of the quantum spacetime geometry as a function of the diffusion time $\sigma$ for three ensembles of causal triangulations characterized by $T=64$ and $k_{0}=1.0$. \subref{FSSspec} Finite size scaling analysis of the spectral homogeneity measure $\mathcal{H}_{S}$ of the quantum spacetime geometry as a function of the diffusion time $\sigma$ for three ensembles of causal triangulations characterized by $T=64$ and $k_{0}=1.0$.}
\label{spechomogeneityFSS}
\end{figure}

In figure \ref{spechomogeneity} I display measurements of the spectral homogeneity measure $\mathcal{H}_{S}(\sigma)$ and, as a reference, the spectral dimension $D_{S}(\sigma)$ for three ensembles of causal triangulations at fixed number $T$ and fixed coupling $k_{0}$. 
For ease of comparison, in figure \ref{spechomogeneityFSS}\subref{allspec} I display all three of these measurements together. In contrast to the volumetric homogeneity $\mathcal{H}_{V}(r)$, the spectral homogeneity measure $\mathcal{H}_{S}(\sigma)$ hardly changes with increasing numbers $N_{3}$ of $3$-simplices. I return to this observation below.

The shape of the homogeneity measure $\mathcal{H}_{S}(\sigma)$ is also readily explained. The homogeneity measure $\mathcal{H}_{S}(\sigma)$ vanishes exactly at $\sigma=0$ by definition and at $\sigma=1$ because all $3$-simplices have four nearest neighbor $3$-simplices. For sufficiently large diffusion times one expects the homogeneity measure $\mathcal{H}_{S}(\sigma)$ to vanish almost exactly because random walks from any starting $3$-simplex now probe the entire causal triangulation. Continuity then essentially dictates the shape of the homogeneity measure $\mathcal{H}_{S}(\sigma)$. The spectral homogeneity measure $\mathcal{H}_{S}(\sigma)$ appears to rise much more rapidly than does the volumetric homogeneity measure $\mathcal{H}_{V}(r)$; however, the former's diffusion time scale is not straightforwardly comparable to the latter's graph geodesic distance scale. I also return to this observation below.

The spectral dimension $D_{S}(\sigma)$ exhibits the well-known phenomenon of dynamical dimensional reduction on sufficiently small diffusion times \cite{JA&JJ&RL7,JA&JJ&RL6,DB&JH,RK}. Interestingly, the homogeneity measure $\mathcal{H}_{S}(\sigma)$ indicates that the quantum spacetime geometry is already extremely homogeneous for diffusion times $\sigma$ less than the diffusion time $\sigma_{\mathrm{max}}$ at which the spectral dimension $D_{S}(\sigma)$ attains its maximum value $D_{S}(\sigma_{\mathrm{max}})$ of approximately the topological dimension of $3$. This finding suggests that inhomogeneity of the quantum spacetime geometry is unrelated to dynamical dimensional reduction. 

In figure \ref{spechomogeneityFSS}\subref{FSSspec} I display the results of a finite size scaling analysis of the spectral homogeneity measure $\mathcal{H}_{S}(\sigma)$ for the same three ensembles of causal triangulations. 
The homogeneity measure $\mathcal{H}_{S}(\sigma)$ is dimensionless, so one does not expect it to finite size scale. The diffusion time is associated with the dimensions of length squared, so one expects it to finite size scale as $\sigma/N_{3}^{2/3}$. The finite size scaled diffusion times $\sigma_{\mathrm{infl}}/N_{3}^{2/3}$, defined below, for the three ensembles are not approximately equivalent, as quantified in figure \ref{spechomogeneityFSS2}\subref{spechomogeneityscalevsN3FSS} below, which suggests that this finite size scaling is not appropriate. 
The plot of figure \ref{spechomogeneityFSS}\subref{allspec}, showing almost no change in the homogeneity measure $\mathcal{H}_{S}(\sigma)$ for increasing numbers $N_{3}$ of $3$-simplices, already hinted at this conclusion. 

\begin{figure}[h!]
\centering
\subfigure[ ]{
\includegraphics[scale=1]{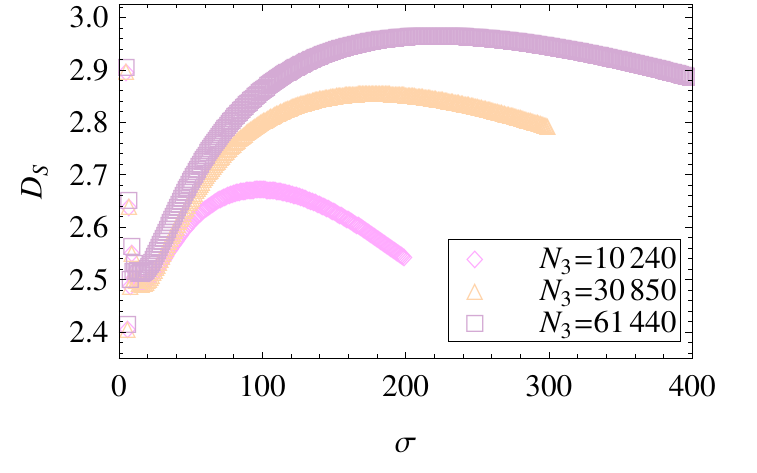}
\label{allspecdim}
}
\subfigure[ ]{
\includegraphics[scale=1]{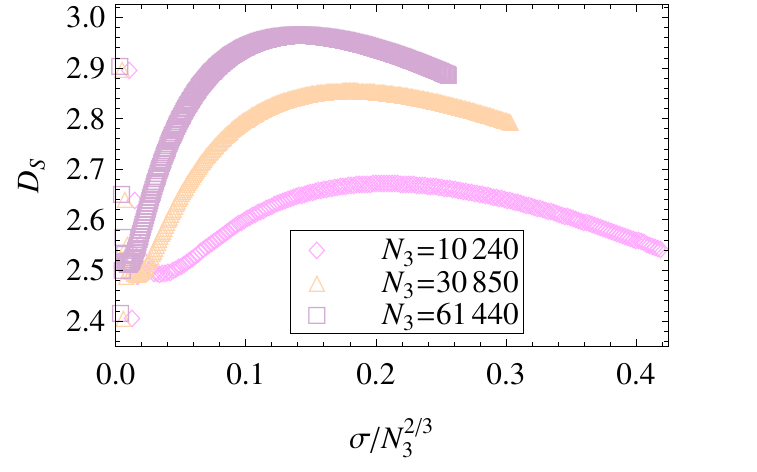}
\label{FSSspecdim}
}
\caption[Optional caption for list of figures]{\subref{allspec} Spectral dimension $D_{S}$ of the quantum spacetime geometry as a function of the diffusion time $\sigma$ for three ensembles of causal triangulations characterized by $T=64$ and $k_{0}=1.0$. \subref{FSSspec} Finite size scaling analysis of the spectral dimension $D_{S}$ of the quantum spacetime geometry as a function of the diffusion time $\sigma$ for three ensembles of causal triangulations characterized by $T=64$ and $k_{0}=1.0$.}
\label{specdimFSS}
\end{figure}

To explore further the appropriateness of the finite size scaling \emph{Ansatz} \eqref{FSSansatz}, 
I display in figure \ref{specdimFSS}\subref{FSSspecdim} the results of a finite size scaling analysis of the spectral dimension $D_{S}(\sigma)$ for the same three ensembles of causal triangulations at fixed number $T$ and fixed coupling $k_{0}$. 
The spectral dimension $D_{S}(\sigma)$ is dimensionless, so one does not expect it to finite size scale. The finite size scaled diffusion times $\sigma_{\mathrm{max}}/N_{3}^{2/3}$ for the three ensembles are also not approximately equivalent, providing further evidence that this finite size scaling is not appropriate. Either of the plots of figure \ref{specdimFSS} demonstrates nonetheless that the suppression of the spectral dimension $D_{S}(\sigma_{\mathrm{max}})$ below the topological value of $3$ for relatively small $N_{3}$ is a finite size effect. 

One might have anticipated that the finite size scaling \emph{Ansatz} \eqref{FSSansatz} is not appropriate for the diffusion times on which I have measured the homogeneity measure $\mathcal{H}_{S}(\sigma)$. 
Whereas Ambj\o rn \emph{et al} justified the finite size scaling \emph{Ansatz} \eqref{FSSansatz} on the basis of the scaling properties of large scale discrete observables admitting semiclassical descriptions \cite{JA&JJ&RL6}, the measurements of the homogeneity measure $\mathcal{H}_{S}(\sigma)$ displayed in figure \ref{spechomogeneity} likely only probe scales on which the quantum geometry is far from semiclassical. Moreover, as I observed above, the homogeneity measure $\mathcal{H}_{S}(\sigma)$ indicates the presence of inhomogeneity only on diffusion times for which the spectral dimension $D_{S}(\sigma)$ is significantly reduced. 
Benedetti and Henson found that the shape of the spectral dimension $D_{S}(\sigma)$ begins to match that of Euclidean de Sitter space only for diffusion times somewhat larger $\sigma_{\mathrm{max}}$ \cite{DB&JH}. 

These considerations inform a qualitative comparison of the graph geodesic distances $r$ on which I have measured the volumetric homogeneity measure $\mathcal{H}_{V}(r)$ and the diffusion times $\sigma$ on which I have measured the spectral homogeneity measure $\mathcal{H}_{S}(\sigma)$. 
The analysis of subsubsection \ref{spacetimehomoresultsvol} indicates that the volumetric homogeneity measure finite size scales canonically, and the analysis of this section indicates that the spectral homogeneity measure finite size scales anomalously. 
These findings suggest that one unit of graph geodesic distance is much larger than one unit of diffusion time. One could attempt to estimate directly the equivalent graph geodesic distance of one diffusion time step by considering the typical graph geodesic distance traversed in one step of the random walk. 
If this suggestion holds true, then the two homogeneity measures do not evince inhomogeneity on the same physical scale and do not signal the transition to homogeneity on the same physical scale. 

I again wish to quantify the typical scale, as measured by the diffusion time $\sigma$, on which the quantum spacetime geometry exhibits inhomogeneity and the typical magnitude, as measured by the homogeneity measure $\mathcal{H}_{S}(\sigma)$, of inhomogeneity of the quantum spacetime geometry on this typical scale. Given the shape of the homogeneity measure $\mathcal{H}_{S}(\sigma)$,  
I take the diffusion time $\sigma_{\mathrm{infl}}$ at the inflection point of the homogeneity measure $\mathcal{H}_{S}(\sigma)$ as the typical scale of inhomogeneity and 
the homogeneity measure $\mathcal{H}_{S}(\sigma_{\mathrm{infl}})$ at the diffusion time $\sigma_{\mathrm{infl}}$ as the typical magnitude of inhomogeneity. I choose the inflection point not the maximum of the homogeneity measure $\mathcal{H}_{S}(\sigma)$ because the maximum occurs at a diffusion time for which random walks are definitely still probing the discreteness of the quantum geometry. I now consider three particular aspects of the finite size scaling analysis. 
In figure \ref{spechomogeneityFSS2}\subref{spechomogeneityscalevsN3} I display the diffusion time $\sigma_{\mathrm{infl}}$ as a function of the number $N_{3}$ of $3$-simplices, which evidently exhibits a constant relation. 
\begin{figure}[h!]
\centering
\subfigure[ ]{
\includegraphics[scale=0.59]{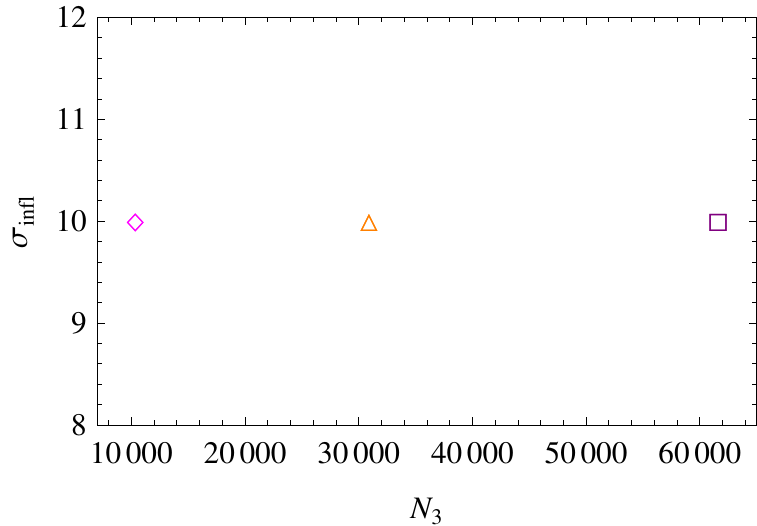}
\label{spechomogeneityscalevsN3}
}
\subfigure[ ]{
\includegraphics[scale=0.64]{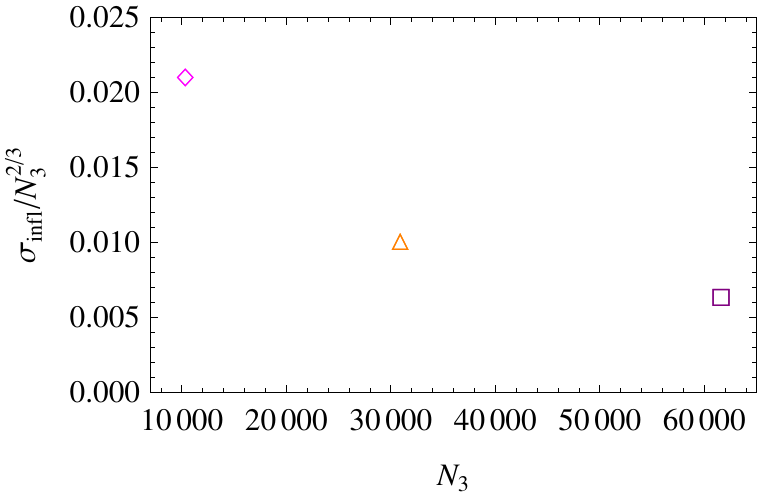}
\label{spechomogeneityscalevsN3FSS}
}
\subfigure[ ]{
\includegraphics[scale=0.7]{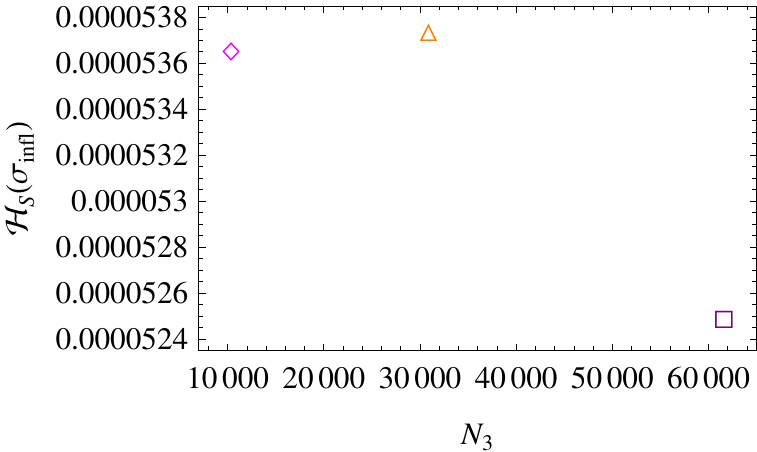}
\label{spechomogeneitymagvsN3}
}
\caption{\subref{spechomogeneityscalevsN3} The diffusion time $\sigma_{\mathrm{infl}}$ as a function of the number $N_{3}$ of $3$-simplices for three ensembles of causal triangulations characterized by $T=64$ and $k_{0}=1.0$. \subref{spechomogeneityscalevsN3FSS} The finite size scaled diffusion time $\sigma_{\mathrm{infl}}/N_{3}^{2/3}$ as a function of the number $N_{3}$ of $3$-simplices for three ensembles of causal triangulations characterized by $T=64$ and $k_{0}=1.0$. \subref{spechomogeneitymagvsN3} The spectral homogeneity measure $\mathcal{H}_{S}(r\sigma_{\mathrm{infl}})$ of the quantum spacetime geometry as a function of the number $N_{3}$ of $3$-simplices for three ensembles of causal triangulations characterized by $T=64$ and $k_{0}=1.0$.}
\label{spechomogeneityFSS2}
\end{figure}
In figure \ref{spechomogeneityFSS2}\subref{spechomogeneityscalevsN3FSS} I display the finite size scaled graph geodesic distance $\sigma_{\mathrm{infl}}/N_{3}^{2/3}$ as a function of the number $N_{3}$ of $3$-simplices, which attests to an inverse relation. As I explained above, these two findings point to the inappropriateness of the finite size scaling. In figure \ref{spechomogeneityFSS2}\subref{spechomogeneitymagvsN3} I display the homogeneity measure $\mathcal{H}_{S}(\sigma_{\mathrm{infl}})$ as a function of the number $N_{3}$ of $3$-simplices, which preliminarily suggests an approximately constant relation. 


Since the finite size scaling \emph{Ansatz} \eqref{FSSansatz} fails to apply to the homogeneity measure $\mathcal{H}_{S}(\sigma)$ for the diffusion times considered, what can one to conclude concerning the limit of arbitrarily large $N_{3}$? Reasoning from the plot of figure \ref{specdimFSS}\subref{allspecdim}, one simply concludes that a finite amount of inhomogeneity persists in this limit. 

\subsection{Temporal evolution of spatial homogeneity}\label{spatialhomoresults}

\begin{figure}[h!]
\centering
\includegraphics[scale=1]{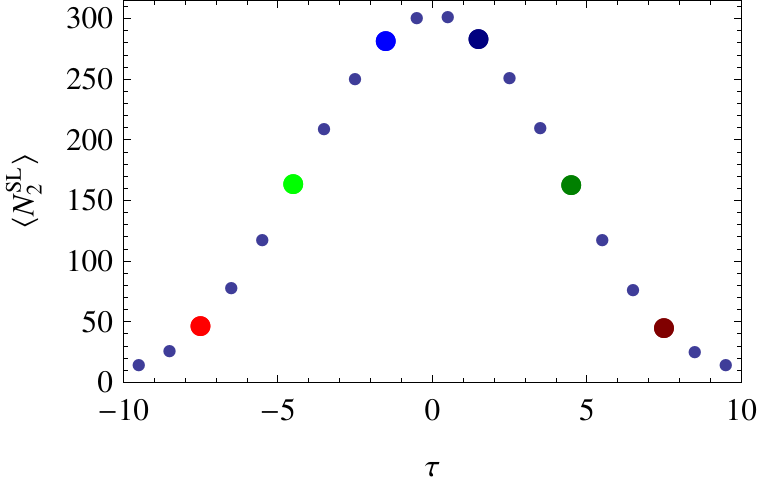}
\caption[Optional caption for list of figures]{Ensemble average number $\langle N_{2}^{\mathrm{SL}}\rangle$ of spacelike $2$-simplices within the central accumulation as a function of the discrete time coordinate $\tau$ for an ensemble of causal triangulations characterized by $T=64$, $N_{3}=10850$, $k_{0}=1.0$. The colors red, green, blue, dark blue, dark green, and dark red correspond respectively to the discrete time coordinate values $\tau=-15/2$, $\tau=-9/2$, $\tau=-3/2$, $\tau=+3/2$, $\tau=+9/2$, and $\tau=+15/2$.}
\label{volprofcolored}
\end{figure}

I have measured the temporal evolution of the homogeneity of the quantum spatial geometry for one ensemble of causal triangulations. Specifically, I consider the six sequential leaves of the distinguished foliation within the central accumulation corresponding to the discrete time coordinate values $\tau=-15/2$, $\tau=-9/2$, $\tau=-3/2$, $\tau=+3/2$, $\tau=+9/2$, and $\tau=+15/2$. As a point of reference, I plot in figure \ref{volprofcolored} the ensemble average number $\langle N_{2}^{\mathrm{SL}}(\tau)\rangle$ of spacelike $2$-simplices as a function of the discrete time coordinate $\tau$ for this ensemble.
I highlight these six leaves 
with the respective colors red, green, blue, dark blue, dark green, and dark red.

I estimate the homogeneity measures $H_{V}(r)$ and $H_{S}(\sigma)$ as explained in appendix \ref{appendix}. 
I first display the results of the measurements for each leaf, and I then consider how the typical scale on which inhomogeneity occurs and the typical magnitude of inhomogeneity on this scale vary with the leaf's discrete spatial $2$-volume. The discrete spatial $2$-volume encodes the history of expansion (and contraction) of the quantum geometry. As in the standard cosmological model of our own universe, the discrete spatial $2$-volume thus serves as a physical proxy for the relevant cosmological time coordinate. I look in particular for power-law scaling of either the typical scale of inhomogeneity or the typical magnitude of inhomogeneity with the discrete spatial $2$-volume. The ubiquity of power-law scalings in the standard cosmological model motivates this analysis. 
For instance, the cosmological evolutions of the various components of our universe---radiation, matter, dark matter, and dark energy---follow power-laws in the scale factor, and cosmic inflation generically predicts that the power spectrum of perturbations propagating on the FLRW spacetime is a power-law in the comoving wavenumber \cite{EWK&MT}. 

\subsubsection{Volumetric measure}

\begin{figure}[h!]
\centering
\subfigure[ ]{
\includegraphics[scale=0.875]{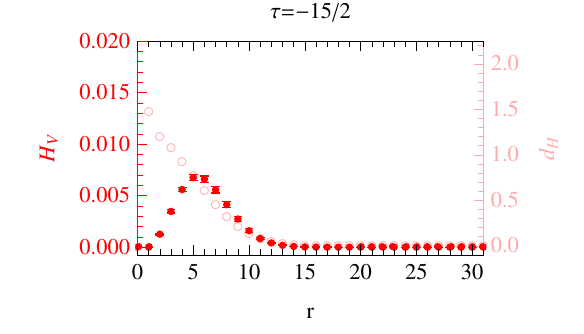}
\label{spatialvolhomogeneity1}
}
\subfigure[ ]{
\includegraphics[scale=0.875]{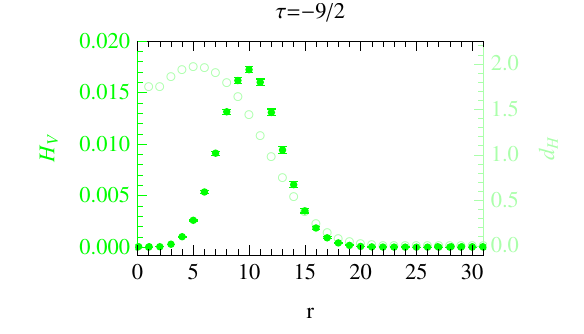}
\label{spatialvolhomogeneity2}
}
\subfigure[ ]{
\includegraphics[scale=0.875]{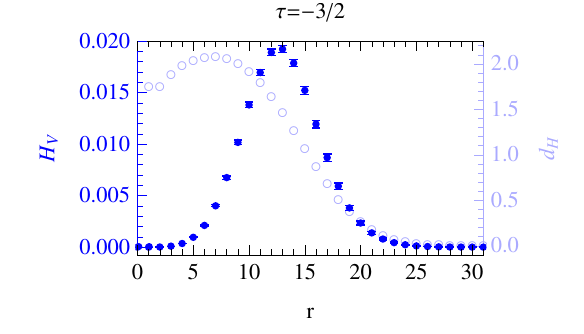}
\label{spatialvolhomogeneity3}
}
\subfigure[ ]{
\includegraphics[scale=0.875]{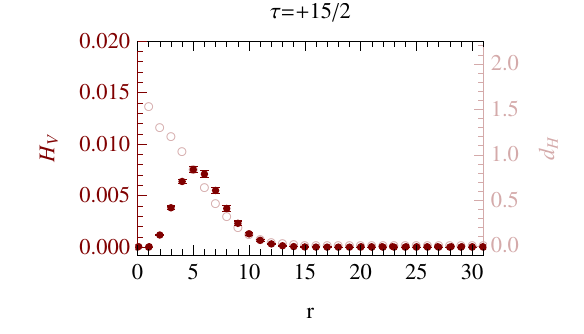}
\label{spatialvolhomogeneity6}
}
\subfigure[ ]{
\includegraphics[scale=0.875]{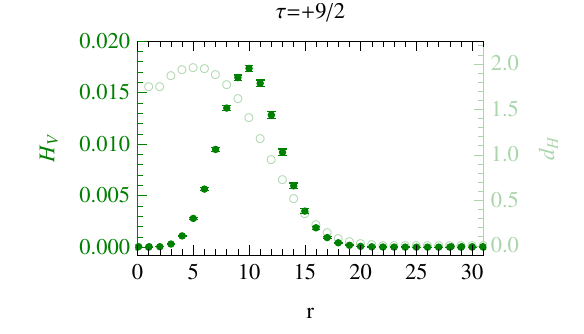}
\label{spatialvolhomogeneity5}
}
\subfigure[ ]{
\includegraphics[scale=0.875]{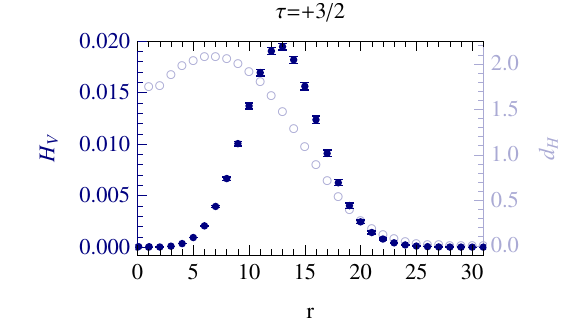}
\label{spatialvolhomogeneity4}
}

\caption[Optional caption for list of figures]{Volumetric homogeneity measure $H_{V}$ (dark, left axis) and Hausdorff dimension $d_{\mathrm{H}}$ (light, right axis) of the quantum spatial geometry as a function of the graph geodesic distance $r$ for an ensemble of causal triangulations characterized by $T=64$, $N_{3}=10850$, $k_{0}=1.0$. \subref{spatialvolhomogeneity1} Leaf of the distinguished foliation at $\tau=-15/2$ (red). \subref{spatialvolhomogeneity2} Leaf of the distinguished foliation at $\tau=-9/2$ (green). \subref{spatialvolhomogeneity3} Leaf of the distinguished foliation at $\tau=-3/2$ (blue). \subref{spatialvolhomogeneity6} Leaf of the distinguished foliation at $\tau=+15/2$ (dark red). \subref{spatialvolhomogeneity5} Leaf of the distinguished foliation at $\tau=+9/2$ (dark green). \subref{spatialvolhomogeneity4} Leaf of the distinguished foliation at $\tau=+3/2$ (dark blue).}
\label{spatialvolhomogeneity}
\end{figure}

\begin{figure}[h!]
\centering
\includegraphics[scale=1]{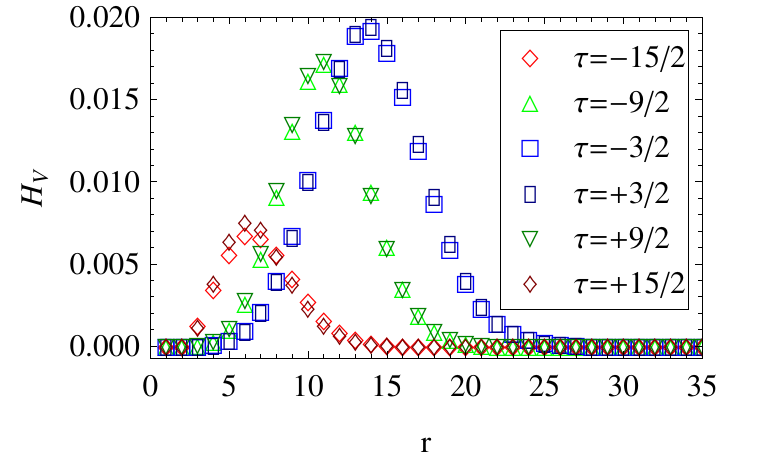}
\caption[Optional caption for list of figures]{Volumetric homogeneity measure $H_{V}$ of the quantum spatial geometry as a function of the graph geodesic distance $r$ for an ensemble of causal triangulations characterized by $T=64$, $N_{3}=10850$, $k_{0}=1.0$.}
\label{spatialvolhomogeneityall}
\end{figure}

In figure \ref{spatialvolhomogeneity} I display measurements of the volumetric homogeneity measure $H_{V}(r)$ and, as a reference, the Hausdorff dimension $d_{\mathrm{H}}(r)$ for six sequential leaves of the distinguished foliation. 
For ease of comparison, in figure \ref{spatialvolhomogeneityall} I display all six of these measurements together. 
The homogeneity measure $H_{V}(r)$ has the same shape as the homogeneity measure $\mathcal{H}_{V}(r)$ for the reason explained in subsubsection \ref{spacetimehomoresultsvol}. A comparison of the measurements from the six leaves yields the following clear patterns. The graph geodesic distance $r_{\mathrm{max}}$ and the homogeneity measure $H_{V}(r_{\mathrm{max}})$, both regarded as functions of the discrete time coordinate $\tau$, increase 
for $-15/2<\tau<0$ and decrease 
for $0<\tau<+15/2$. 
Accordingly, the graph geodesic distance $r_{\mathrm{max}}$ and the homogeneity measure $H_{V}(r_{\mathrm{max}})$ track the temporal evolution of the discrete spatial $2$-volume as measured by the ensemble average number $\langle N_{2}^{\mathrm{SL}}\rangle$ of spacelike $2$-simplices. 
Generally, the graph geodesic distance $r_{\mathrm{max}}$ and the homogeneity measure $H_{V}(r_{\mathrm{max}})$ also track the temporal evolution of the uncertainty in the discrete spatial $2$-volume as measured by the diagonal of the ensemble average covariance $\langle n_{2}^{\mathrm{SL}}(\tau)n_{2}^{\mathrm{SL}}(\tau')\rangle$. To determine how closely the graph geodesic distance $r_{\mathrm{max}}$ and the homogeneity measure $H_{V}(r_{\mathrm{max}})$ correlate with this uncertainty requires measurements of the homogeneity measure $H_{V}(r)$ for most of the twenty leaves of the distinguished foliation.

\begin{figure}[h!]
\centering
\subfigure[ ]{
\includegraphics[scale=0.9]{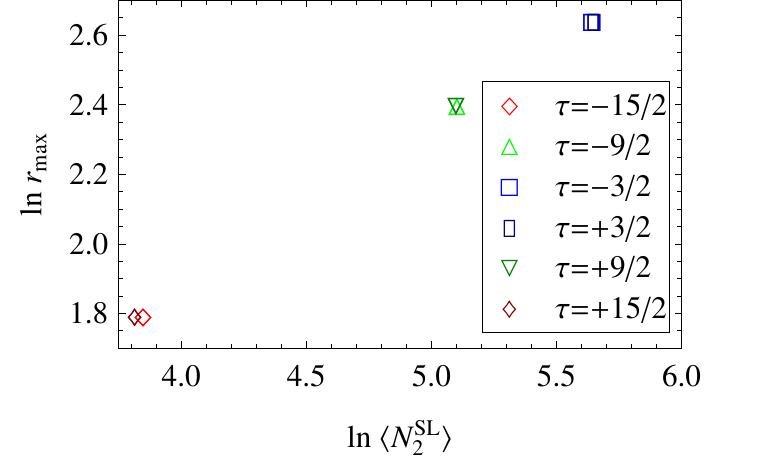}
\label{volhomoscalevs2vol}
}
\subfigure[ ]{
\includegraphics[scale=0.75]{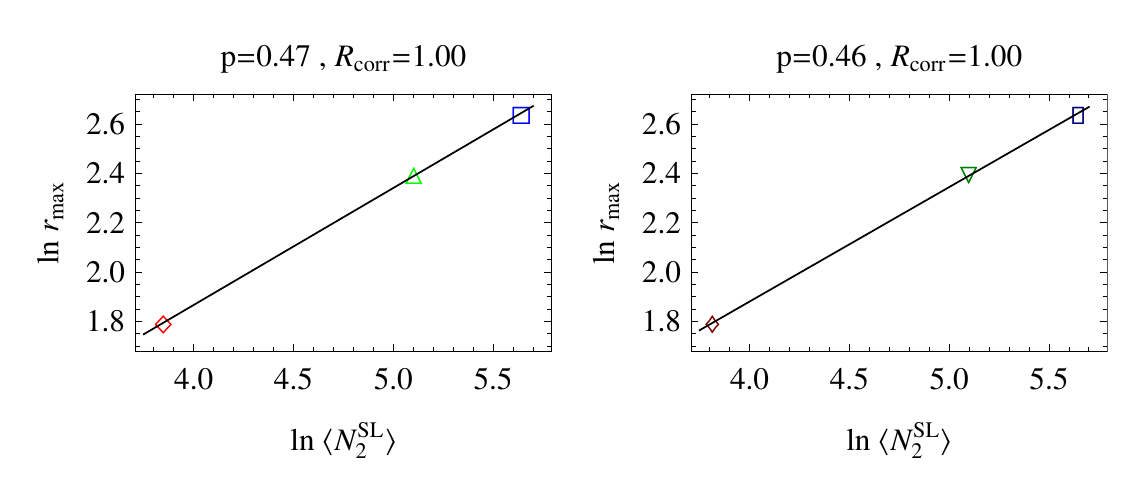}
\label{volhomoscalevs2volfit}
}
\caption[Optional caption for list of figures]{\subref{volhomoscalevs2vol} Logarithm of the graph geodesic distance $r_{\mathrm{max}}$ as a function of the logarithm of the ensemble average number $\langle N_{2}^{\mathrm{SL}}\rangle$ of spacelike $2$-simplices for an ensemble of causal triangulations characterized by $T=64$, $N_{3}=10850$, $k_{0}=1.0$. \subref{volhomoscalevs2volfit} Linear fit for the discrete time coordinate values $\tau=-15/2$, $\tau=-9/2$, and $\tau=-3/2$ (left) and linear fit for the discrete time coordinate values $\tau=+3/2$, $\tau=+9/2$, and $\tau=+15/2$ (right).}
\label{volhomoscale}
\end{figure}

I now quantify these trends, looking specifically for power-law scaling of the graph geodesic distance $r_{\mathrm{max}}$ and the homogeneity measure $H_{V}(r_{\mathrm{max}})$ with the ensemble average $\langle N_{2}^{\mathrm{SL}}\rangle$. 
In figure \ref{volhomoscale}\subref{volhomoscalevs2vol} I plot the logarithm of the graph geodesic distance $r_{\mathrm{max}}$ as a function of the logarithm of the ensemble average $\langle N_{2}^{\mathrm{SL}}\rangle$. 
In figure \ref{volhomoscale}\subref{volhomoscalevs2volfit} I plot a linear fit to the logarithm of the graph geodesic distance $r_{\mathrm{max}}$ as a function of the logarithm of the ensemble average $\langle N_{2}^{\mathrm{SL}}\rangle$ for the discrete time coordinate values $\tau=-15/2$, $\tau=-9/2$, and $\tau=-3/2$ corresponding to increasing discrete spatial $2$-volume and 
for the discrete time coordinate values $\tau=+3/2$, $\tau=+9/2$, and $\tau=+15/2$ corresponding to decreasing discrete spatial $2$-volume. These two linear fits provide strong evidence for power-law scaling of the graph geodesic distance $r_{\mathrm{max}}$ with the ensemble average $\langle N_{2}^{\mathrm{SL}}\rangle$ for a power of nearly $1/2$.

\begin{figure}[h!]
\centering
\subfigure[ ]{
\includegraphics[scale=0.9]{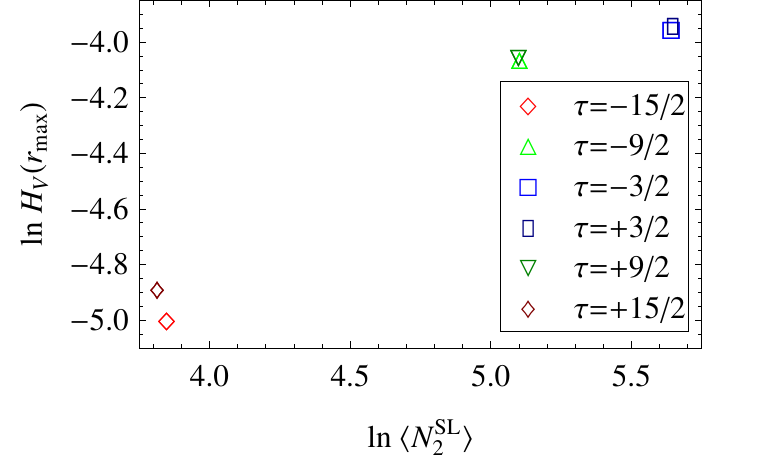}
\label{volhomomagvs2vol}
}
\subfigure[ ]{
\includegraphics[scale=0.75]{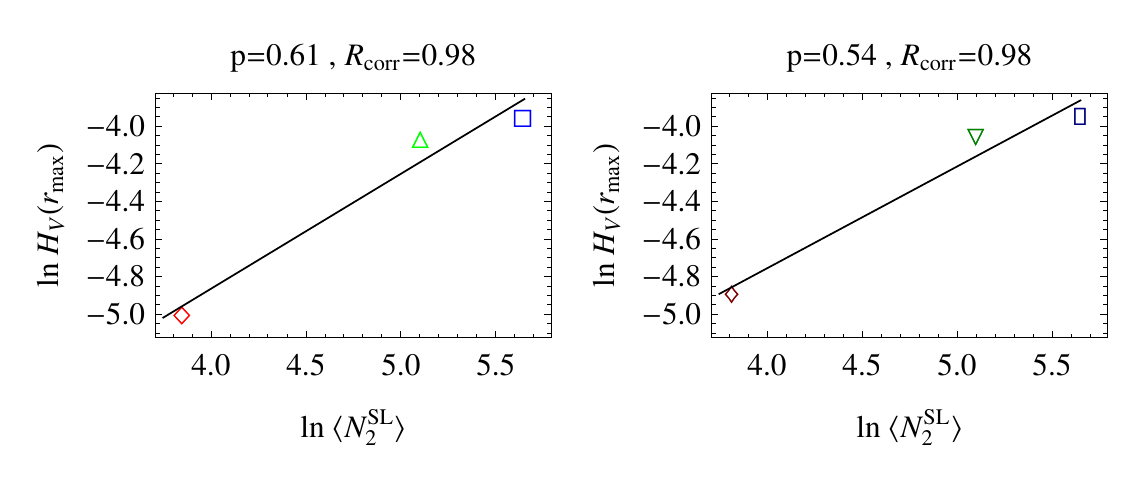}
\label{volhomomagvs2volfit}
}
\caption[Optional caption for list of figures]{\subref{volhomomagvs2vol} Logarithm of the homogeneity measure $H_{V}(r_{\mathrm{max}})$ as a function of the logarithm of the ensemble average number $\langle N_{2}^{\mathrm{SL}}(\tau)\rangle$ of spacelike $2$-simplices for an ensemble of causal triangulations characterized by $T=64$, $N_{3}=10850$, $k_{0}=1.0$. \subref{volhomomagvs2volfit} Linear fit for the discrete time coordinate values $\tau=-15/2$, $\tau=-9/2$, and $\tau=-3/2$ (left) and linear fit for the discrete time coordinate values $\tau=+3/2$, $\tau=+9/2$, and $\tau=+15/2$ (right).}
\label{volhomomag}
\end{figure}

In figure \ref{volhomomag}\subref{volhomomagvs2vol} I plot the logarithm of the homogeneity measure $H_{V}(r_{\mathrm{max}})$ as a function of the logarithm of the ensemble average $\langle N_{2}^{\mathrm{SL}}\rangle$. 
In figure \ref{volhomomag}\subref{volhomomagvs2volfit} I plot a linear fit to the logarithm of the homogeneity measure $H_{V}(r_{\mathrm{max}})$ as a function of the logarithm of the ensemble average $\langle N_{2}^{\mathrm{SL}}\rangle$ for the discrete time coordinate values $\tau=-15/2$, $\tau=-9/2$, and $\tau=-3/2$ corresponding to increasing discrete spatial $2$-volume and 
for the discrete time coordinate values $\tau=+3/2$, $\tau=+9/2$, and $\tau=+15/2$ corresponding to decreasing discrete spatial $2$-volume. These two linear fits provide evidence for subleading power-law scaling of the homogeneity measure $H_{V}(r_{\mathrm{max}})$ with the ensemble average $\langle N_{2}^{\mathrm{SL}}\rangle$ for a power of approximately $0.57$.

\subsubsection{Spectral measure}

\begin{figure}[h!]
\centering
\subfigure[ ]{
\includegraphics[scale=0.875]{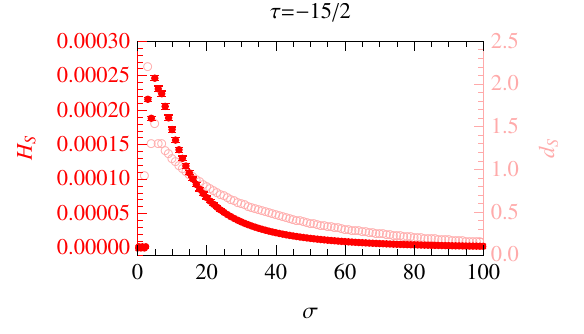}
\label{spatialspechomogeneity1}
}
\subfigure[ ]{
\includegraphics[scale=0.875]{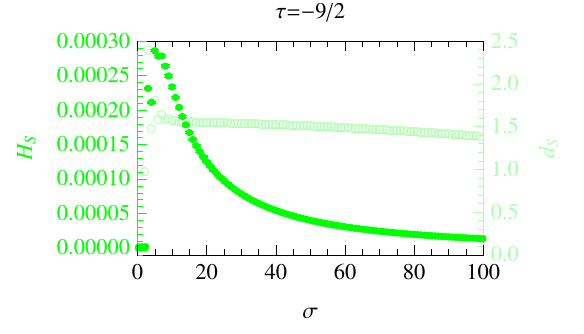}
\label{spatialspechomogeneity2}
}
\subfigure[ ]{
\includegraphics[scale=0.875]{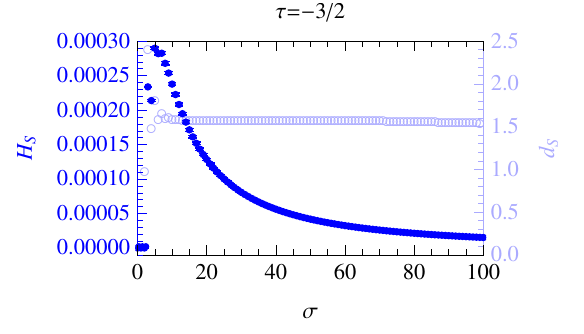}
\label{spatialspechomogeneity3}
}
\subfigure[ ]{
\includegraphics[scale=0.875]{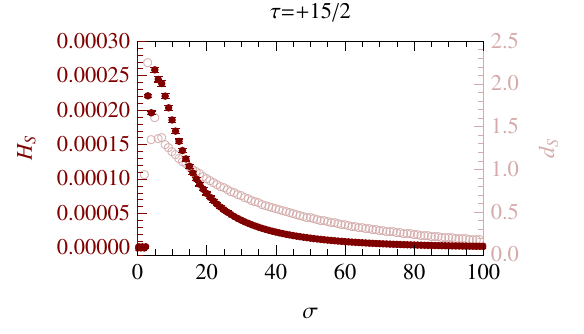}
\label{spatialspechomogeneity6}
}
\subfigure[ ]{
\includegraphics[scale=0.875]{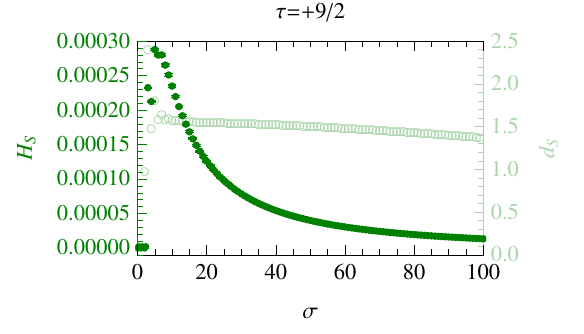}
\label{spatialspechomogeneity5}
}
\subfigure[ ]{
\includegraphics[scale=0.875]{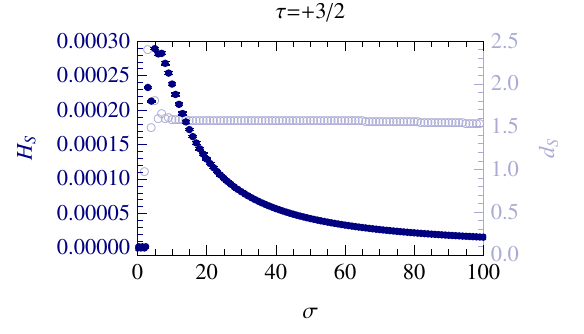}
\label{spatialspechomogeneity4}
}
\caption[Optional caption for list of figures]{Spectral homogeneity measure $H_{S}$ (dark, left axis) and spectral dimension $d_{\mathrm{S}}$ (light, right axis) of the quantum spatial geometry as a function of the diffusion time $\sigma$ for an ensemble of causal triangulations characterized by $T=64$, $N_{3}=10850$, $k_{0}=1.0$. \subref{spatialspechomogeneity1} Leaf of the distinguished foliation at $\tau=-15/2$ (red). \subref{spatialspechomogeneity2} Leaf of the distinguished foliation at $\tau=-9/2$ (green). \subref{spatialspechomogeneity3} Leaf of the distinguished foliation at $\tau=-3/2$ (blue). \subref{spatialspechomogeneity6} Leaf of the distinguished foliation at $\tau=+15/2$ (dark red). \subref{spatialspechomogeneity5} Leaf of the distinguished foliation at $\tau=+9/2$ (dark green). \subref{spatialspechomogeneity4} Leaf of the distinguished foliation at $\tau=+3/2$ (dark blue).}
\label{spatialspechomogeneity}
\end{figure}

\begin{figure}[h!]
\centering
\includegraphics[scale=1]{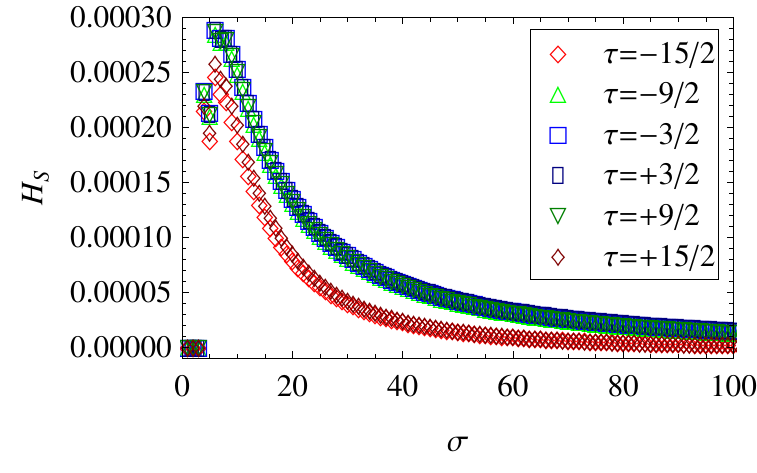}
\caption[Optional caption for list of figures]{Spectral homogeneity measure $H_{S}$ of the quantum spatial geometry as a function of the diffusion time $\sigma$ for an ensemble of causal triangulations characterized by $T=64$, $N_{3}=10850$, $k_{0}=1.0$.}
\label{spatialspechomogeneityall}
\end{figure}

In figure \ref{spatialspechomogeneity} I display measurements of the spectral homogeneity measure $H_{S}(\sigma)$ and, as a reference, the spectral dimension $d_{S}(\sigma)$ for six sequential leaves of the distinguished foliation.
For ease of comparison, in figure \ref{spatialspechomogeneityall} I display all six of these measurements together. 
The homogeneity measure $H_{S}(\sigma)$ for all six quantum spatial geometries has the same shape as the homogeneity measure $\mathcal{H}_{S}(\sigma)$ for the reason explained in subsubsection \ref{spacetimehomoresultsspec}. The measurements of the spectral dimension $d_{S}(\sigma)$ for the four leaves labeled by the discrete time coordinate values $\tau=-9/2$, $\tau=-3/2$, $\tau=+3/2$, and $\tau=+9/2$, showing a constant value of approximately $3/2$, are consistent with previous result \cite{JA&JJ&RL6,RK}. Presumably, the measurements of the spectral dimension $d_{S}(\sigma)$ for the two leaves labeled by the discrete time coordinate value $\tau=-15/2$ and $\tau=+15/2$, showing a falloff from an initial value of approximately $3/2$, are dominated by the significant positive curvature of the contributing spacelike $2$-surfaces. 

A comparison of the measurements from the six leaves yields the following clear patterns. The diffusion time $\sigma_{\mathrm{infl}}$ and the homogeneity measure $H_{S}(\sigma_{\mathrm{infl}})$, both regarded as functions of the discrete time coordinate $\tau$, increase for $-15/2<\tau<0$ and decrease for $0<\tau<+15/2$. Accordingly, the diffusion time $\sigma_{\mathrm{infl}}$ and the homogeneity measure $H_{S}(\sigma_{\mathrm{infl}})$ track the temporal evolution of the discrete spatial $2$-volume as measured by the ensemble average number $\langle N_{2}^{\mathrm{SL}}\rangle$ of spacelike $2$-simplices. 
Generally, the diffusion time $\sigma_{\mathrm{infl}}$ and the homogeneity measure $H_{S}(\sigma_{\mathrm{infl}})$ also track the temporal evolution of the uncertainty in the discrete spatial $2$-volume as measured by the diagonal of the ensemble average covariance $\langle n_{2}^{\mathrm{SL}}(\tau)n_{2}^{\mathrm{SL}}(\tau')\rangle$. To determine how closely the diffusion time $\sigma_{\mathrm{infl}}$ and the homogeneity measure $H_{S}(\sigma_{\mathrm{infl}})$ correlate with this uncertainty requires measurements of the homogeneity measure $H_{S}(\sigma)$ for most of the twenty leaves of the distinguished foliation.

\begin{figure}[h!]
\centering
\includegraphics[scale=0.9]{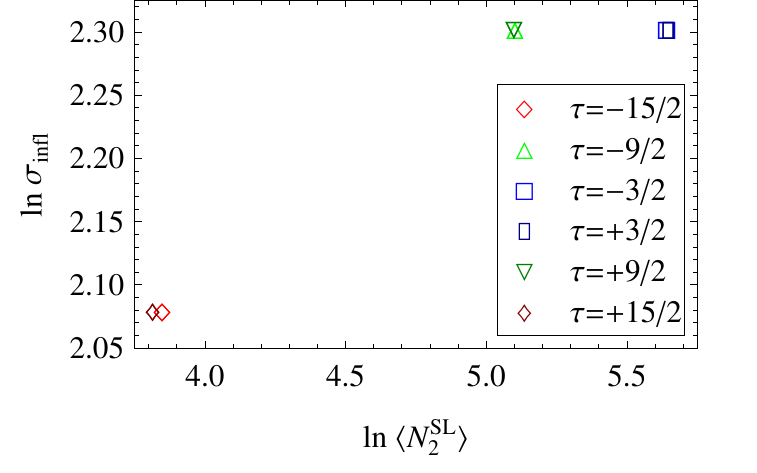}
\caption[Optional caption for list of figures]{Logarithm of the diffusion time $\sigma_{\mathrm{infl}}$ as a function of the logarithm of the ensemble average number $\langle N_{2}^{\mathrm{SL}}\rangle$ of spacelike $2$-simplices for an ensemble of causal triangulations characterized by $T=64$, $N_{3}=10850$, $k_{0}=1.0$.}
\label{spechomoscale}
\end{figure}

I now quantify these trends, looking specifically for power-law scaling of the diffusion time 
$\sigma_{\mathrm{infl}}$ and the homogeneity measure $H_{S}(\sigma_{\mathrm{infl}})$ with the ensemble average $\langle N_{2}^{\mathrm{SL}}\rangle$. In figure \ref{spechomoscale} 
I plot the logarithm of the diffusion time $\sigma_{\mathrm{infl}}$ as a function of the logarithm of the ensemble average $\langle N_{2}^{\mathrm{SL}}\rangle$. 
There is clearly no evidence even for subleading power-law scaling of the diffusion time $\sigma_{\mathrm{infl}}$ with the ensemble average $\langle N_{2}^{\mathrm{SL}}\rangle$. 

\begin{figure}[h!]
\centering
\subfigure[ ]{
\includegraphics[scale=0.9]{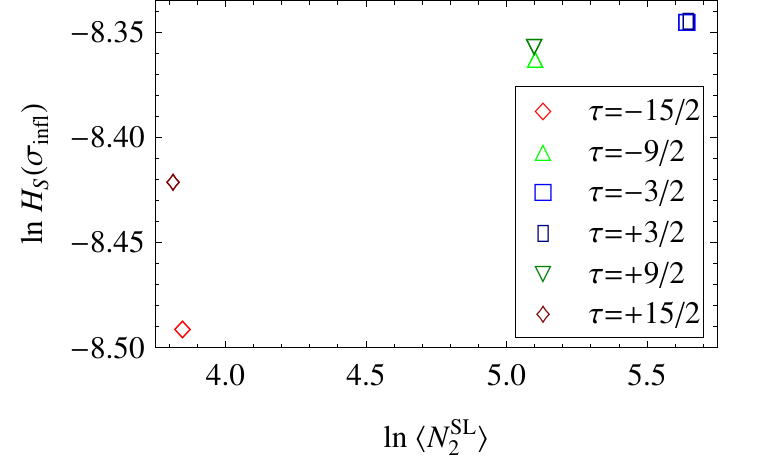}
\label{spechomomagvs2vol}
}
\subfigure[ ]{
\includegraphics[scale=0.75]{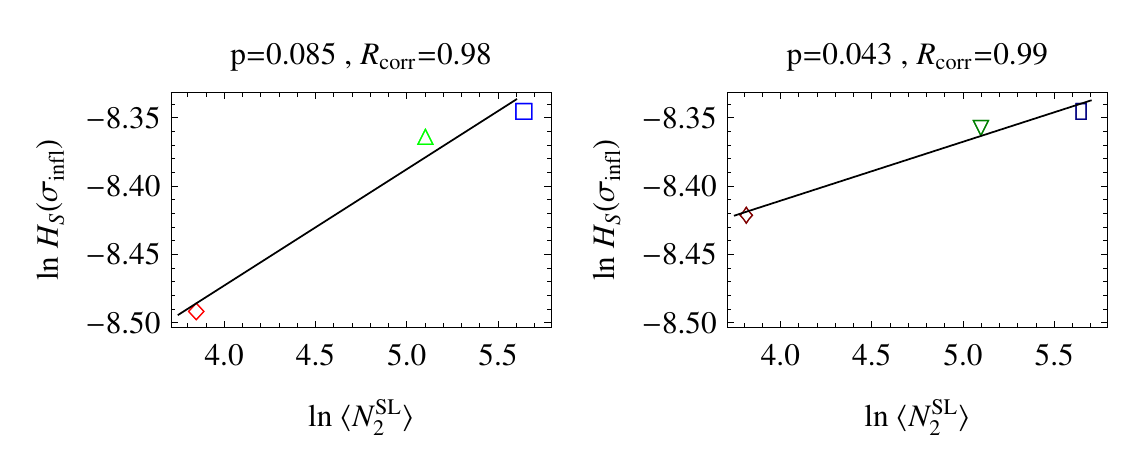}
\label{spechomomagvs2volfit}
}
\caption[Optional caption for list of figures]{\subref{spechomomagvs2vol} Logarithm of the homogeneity measure $H_{V}(\sigma_{\mathrm{infl}})$ as a function of the logarithm of the ensemble average number $\langle N_{2}^{\mathrm{SL}}\rangle$ of spacelike $2$-simplices for an ensemble of causal triangulations characterized by $T=64$, $N_{3}=10850$, $k_{0}=1.0$. \subref{spechomomagvs2volfit} Linear fit for the discrete time coordinate values $\tau=-15/2$, $\tau=-9/2$, and $\tau=-3/2$ corresponding to increasing discrete spatial $2$-volume (left) and for the discrete time coordinate values $\tau=+3/2$, $\tau=+9/2$, and $\tau=+15/2$ corresponding to decreasing discrete spatial $2$-volume (right).}
\label{spechomomag}
\end{figure}

In figure \ref{spechomomag}\subref{spechomomagvs2vol} I plot the logarithm of the homogeneity measure $H_{S}(\sigma_{\mathrm{infl}})$ as a function of the logarithm of the ensemble average $\langle N_{2}^{\mathrm{SL}}\rangle$. 
In figure \ref{spechomomag}\subref{spechomomagvs2volfit} I plot a linear fit to the logarithm of the homogeneity measure $H_{S}(\sigma_{\mathrm{infl}})$ as a function of the logarithm of the ensemble average $\langle N_{2}^{\mathrm{SL}}\rangle$ for the discrete time coordinate values $\tau=-15/2$, $\tau=-9/2$, and $\tau=-3/2$ corresponding to increasing discrete spatial $2$-volume and 
for the discrete time coordinate values $\tau=+3/2$, $\tau=+9/2$, and $\tau=+15/2$ corresponding to decreasing discrete spatial $2$-volume. Each of these linear fits, particularly the latter, provides evidence for at least subleading power-law scaling of the homogeneity measure $H_{S}(\sigma_{\mathrm{infl}})$ with the ensemble average $\langle N_{2}^{\mathrm{SL}}\rangle$. The discrepancy between the homogeneity measure $H_{V}(\sigma_{\mathrm{infl}})$ for $\tau=-15/2$ and for $\tau=+15/2$ calls into question this evidence to some extent.

\section{Conclusion}\label{conclusion}

Within an approach to the construction of quantum theories of fields that employs a lattice regularization, such as causal dynamical triangulations, there are two crucial questions to ask of any discrete observable. Firstly, is the discrete observable well-defined in the continuum limit (assuming that the continuum limit exists)? Subsequently, what continuous observable is the counterpart of the discrete observable in the continuum limit (assuming that the discrete observable is well-defined in the continuum limit)? In the preceding I defined two discrete observables, each a measure of the homogeneity of the quantum geometry determined by an ensemble of causal triangulations, and I studied their properties at finite number of $3$-simplices and finite lattice spacing. 
How do my findings inform these two questions?

The finite size scaling analysis of subsubsection \ref{spacetimehomoresultsvol} indicates that the volumetric homogeneity measure is not well-defined in the limit of arbitrarily many $3$-simplices, 
and the finite size scaling analysis of subsubsection \ref{spacetimehomoresultsspec} indicates that the spectral homogeneity measure does not change appreciably with increasing numbers of $3$-simplices (at least for sufficiently small diffusion times). The latter behavior raises the prospect of the spectral homogeneity measure being well-defined in the limit of arbitrarily many $3$-simplices. The continuum limit involves not only the limit of diverging number of $3$-simplices, but also the limit of vanishing lattice spacing. 
From the measurements of the two homogeneity measures that I have performed so far, 
I simply cannot draw any conclusions about the potential effect of the latter limit. 
Answers to the above two questions must thus await a further study.

This next study should address itself to the $(3+1)$-dimensional quantum theory. As I observed in subsection \ref{phenomenology}, the partition function for the $(2+1)$-dimensional quantum theory exhibits only a first order phase transition whereas the partition function for the $(3+1)$-dimensional quantum theory exhibits also a second order transition. 
Accordingly, the former quantum theory, considered in the preceding, likely does not possess a continuum limit, but the latter quantum theory potentially does possess a continuum limit. 
By studying the two homogeneity measures for ensembles of causal triangulations approaching the second order phase transition, one might gain insight into the above two questions regarding the continuum limit. Ideally, one would consider a succession of ensembles of causal triangulation along a renormalization group trajectory connected to the hypothetical ultraviolet fixed point. Whether or not such renormalization group trajectories exist is currently under investigation \cite{JA&AG&JJ&AK&RL,JHC1}.

Homogeneity is paired with isotropy in the cosmological principle 
as I discussed in section \ref{introduction}. 
Complementing this study of homogeneity, I plan to devise scale-dependent measures of the isotropy of the quantum geometry determined by an ensemble of causal triangulations. Quantifying isotropy is a notably more complicated exercise because isotropy involves not just place, but also direction. Efforts to assess the isotropy of our own universe on the basis of galaxy redshift surveys may again prove insightful \cite{CM&JB&AB}. 


\section*{Acknowledgments}

I thank Christian Anderson, David Kamensky, Jonah Miller, and especially Rajesh Kommu for allowing me to employ parts of their computer codes. I thank Steven Carlip for suggesting the scaling analyses of subsection \ref{spatialhomoresults}. I acknowledge support from the Foundation for Fundamental Research on Matter itself supported by the Netherlands Organization for Scientific Research. 

\appendix

\section{Estimates and errors}\label{appendix}

Let $\mathcal{T}$ be a causal triangulation $\mathcal{T}_{c}$ or a distinguished spacelike $d$-surface $\mathsf{T}_{\tau}$ of a causal triangulation $\mathcal{T}_{c}$. Let $F_{\mathcal{T}}(\ell)$ be the function $\mathfrak{N}_{\mathcal{T}_{c}}(r)$ or $\mathsf{N}_{\mathsf{T}_{\tau}}(r)$ in the volumetric case or the function $P_{\mathcal{T}_{c}}(\sigma)$ or $\mathcal{P}_{\mathsf{T}_{\tau}}(\sigma)$ in the spectral case. Let $N_{D}$ be the number $N_{d+1}$ of $(d+1)$-simplices comprising a causal triangulation or the number $N_{d}$ of $d$-simplices comprising a triangulated spacelike $d$-surface of a causal triangulation. Since the number $N_{D}$ of $D$-simplices is typically quite large, I estimate the variance $\mathrm{var}[F_{\mathcal{T}}(\ell)]$ by considering only a subset of $K$ randomly selected $D$-simplices $s_{k}$:
\begin{equation}
\mathrm{var}[F_{\mathcal{T}}^{(K)}(\ell)]=\frac{1}{K-1}\sum_{s'_{k}\in\mathcal{T}}\left[F_{s'_{k}}(\ell)-\frac{1}{K}\sum_{s_{k}\in\mathcal{T}}F_{s_{k}}(\ell)\right]^{2}.
\end{equation}
One clearly recovers the variance $\mathrm{var}[F_{\mathcal{T}}(\ell)]$ in the limit as $K$ approaches $N_{D}$:
\begin{equation}
\mathrm{var}[F_{\mathcal{T}}(\ell)]=\lim_{K\rightarrow N_{D}}\mathrm{var}[F_{\mathcal{T}}^{(K)}(\ell)].
\end{equation}
I perform a jackknife analysis to estimate the error $\epsilon[\mathrm{var}[F_{\mathcal{T}}^{(K)}(\ell)]]$ in the estimated variance $\mathrm{var}[F_{\mathcal{T}}^{(K)}(\ell)]$ incurred by considering only a subset of $K$ $D$-simplices.\footnote{If $N_{D}$ is not substantially larger than $K$, then I include the appropriate finite population correction factor.} Since the number $N(\mathcal{T})$ of triangulations comprising an ensemble is necessarily finite, I estimate the expectation value $\mathbb{E}[\mathrm{var}[F(\ell)]]$ of the variance $\mathrm{var}[F_{\mathcal{T}}(\ell)]$ by its average over an ensemble:
\begin{equation}
\langle\mathrm{var}[F(\ell)]\rangle=\frac{1}{N(\mathcal{T})}\sum_{l=1}^{N(\mathcal{T})}\mathrm{var}[F_{\mathcal{T}^{(l)}}(\ell)].
\end{equation}
One clearly recovers the expectation value $\mathbb{E}[\mathrm{var}[F(\ell)]]$ in the limit as $N(\mathcal{T})$ diverges without bound:
\begin{equation}
\mathbb{E}[\mathrm{var}[F(\ell)]]=\lim_{N(\mathcal{T})\rightarrow\infty}\langle\mathrm{var}[F(\ell)]\rangle.
\end{equation}
I perform a jackknife analysis to estimate the error $\epsilon[\langle\mathrm{var}[F(\ell)]\rangle]$ in the ensemble average variance $\langle\mathrm{var}[F(\ell)]\rangle$ incurred by considering only a finite ensemble of $N(\mathcal{T})$ triangulations. Now let $\mathsf{H}(\ell)$ be the homogeneity measure $\mathcal{H}_{V}(r)$ or $H_{V}(r)$ in the volumetric case or the homogeneity measure $\mathcal{H}_{S}(\sigma)$ or $H_{S}(\sigma)$ in the spectral case. Taking both of the above estimations into account, I then estimate the homogeneity measure $\mathsf{H}(\ell)$ as
\begin{equation}
\langle\mathsf{H}^{(K)}(\ell)\rangle=\langle\mathrm{var}[F^{(K)}(\ell)]\rangle.
\end{equation}
One clearly recovers the homogeneity measure $\mathsf{H}(\ell)$ in the double limit:
\begin{equation}
\mathsf{H}(\ell)=\lim_{\substack{K\rightarrow N_{D} \\ N(\mathcal{T})\rightarrow\infty}}\langle\mathsf{H}^{(K)}(\ell)\rangle.
\end{equation}
I estimate the error $\epsilon[\langle\mathsf{H}^{(K)}(\ell)\rangle]$ in the estimate $\langle\mathsf{H}^{(K)}(\ell)\rangle$ of the homogeneity measure $\mathsf{H}(\ell)$ as
\begin{equation}
\epsilon[\langle\mathsf{H}^{(K)}(\ell)\rangle]=\langle\epsilon[\mathrm{var}[F^{(K)}(\ell)]]\rangle+\epsilon[\langle\mathrm{var}[F(\ell)]\rangle],
\end{equation}
where
\begin{equation}
\langle\epsilon[\mathrm{var}[F^{(K)}(\ell)]]\rangle=\sqrt{\frac{1}{N^{2}(\mathcal{T})}\sum_{l=1}^{N(\mathcal{T})}\epsilon^{2}[\mathrm{var}[F_{\mathcal{T}^{(l)}}^{(K)}(\ell)]]}
\end{equation}
is the error propagated into the estimated variance $\mathrm{var}[F^{(K)}(\ell)]$ from the $N(\mathcal{T})$ triangulations.


\begin{thebibliography}{99}

\bibitem{BICEP2} P. A. R. Ade \emph{et al}. ``Detection of $B$-Mode Polarization at Degree Angular Scales by BICEP2." \emph{Physical Review Letters} 112 (2014) 241101.

\bibitem{Alonso} D. Alonso, A. Bueno Belloso, F. J. S\'anchez, J. Garc'a-Bellido, and E. S\'anchez ``Measuring the transition to homogeneity with photometric redshift survey." \emph{Monthly Notices of the Royal Astronomical Society} 440 (2014) 10. 

\bibitem{JA&TB} J. Ambj{\o}rn and T. Budd. ``Geodesic distances in quantum Liouville gravity." arXiv: hep-th/1405.3424.

\bibitem{JA&AG&JJ&AK&RL} J. Ambj{\o}rn, A. G\"{o}rlich, J. Jurkiewicz, A. Kreienbuehl, and R. Loll. ``Renormalization group flow in CDT." \emph{Classical and Quantum Gravity} 31 (2014) 165003. 

\bibitem{JA&AG&JJ&RL1} J. Ambj{\o}rn, A. G\"{o}rlich, J. Jurkiewicz, and R. Loll. ``Planckian Birth of a Quantum de Sitter Universe." \emph{Physical Review Letters} 100 (2008) 091304.

\bibitem{JA&AG&JJ&RL2}  J. Ambj{\o}rn, A. G\"{o}rlich, J. Jurkiewicz, and R. Loll. ``Nonperturbative quantum de Sitter universe."  \emph{Physical Review D} 78 (2008) 063544.

\bibitem{JA&AG&JJ&RL3}  J. Ambj{\o}rn, A. G\"{o}rlich, J. Jurkiewicz, and R. Loll. ``Nonperturbative quantum gravity." \emph{Physics Reports} 519 (2012) 127.

\bibitem{JA&AG&JJ&RL&JGS&TT} J. Ambj{\o}rn, A. G\"{o}rlich, J. Jurkiewicz, R. Loll, J. Gizbert-Studnicki, and T. Trze\'{s}niewski. ``The semiclassical limit of causal dynamical triangulations." \emph{Nuclear Physics B} 849 (2011) 144.

\bibitem{JA&SJ&JJ&RL1} J. Ambj{\o}rn, S. Jordan, J. Jurkiewicz, and R. Loll. ``Second-Order Phase Transition in Causal Dynamical Triangulations." \emph{Physical Review Letters} 107 (2011) 211303.

\bibitem{JA&SJ&JJ&RL2} J. Ambj{\o}rn, S. Jordan, J. Jurkiewicz, and R. Loll. ``Second- and first-order phase transitions in causal dynamical triangulations." \emph{Physical Review D} 85 (2012) 124044.

\bibitem{JA&JJ&RL1} J. Ambj{\o}rn, J. Jurkiewicz, and R. Loll. ``Non-perturbative Lorentzian Path Integral for Gravity." \emph{Physical Review Letters} 85 (2000) 347.

\bibitem{JA&JJ&RL2} J. Ambj{\o}rn, J. Jurkiewicz, and R. Loll. ``Dynamically triangulating Lorentzian quantum gravity." \emph{Nuclear Physics B} 610 (2001) 347.

\bibitem{JA&JJ&RL3} J. Ambj{\o}rn, J. Jurkiewicz, and R. Loll. ``Nonperturbative 3d Lorentzian Quantum Gravity." \emph{Physical Review D} 64 (2001) 044011.

\bibitem{JA&JJ&RL4} J. Ambj{\o}rn, J. Jurkiewicz, and R. Loll. ``Emergence of a 4D World from Causal Dynamical Triangulations." \emph{Physical Review Letters} 93 (2004) 131301.

\bibitem{JA&JJ&RL5} J. Ambj{\o}rn, J. Jurkiewicz, and R. Loll. ``Semiclassical universe from first principles." \emph{Physics Letters B} 607 (2005) 205.

\bibitem{JA&JJ&RL7} J. Ambj{\o}rn, J. Jurkiewicz, and R. Loll. ``The Spectral Dimension of the Universe is Scale Dependent."  \emph{Physical Review Letters} 95 (2005) 171301.

\bibitem{JA&JJ&RL6} J. Ambj{\o}rn, J. Jurkiewicz, and R. Loll. ``Reconstructing the universe." \emph{Physical Review D} 72 (2005) 064014.

\bibitem{CA&SJC&JHC&PH&RKK&PZ} C. Anderson, S. J. Carlip, J. H. Cooperman, P. Ho\v{r}ava, R. K. Kommu, and P. R. Zulkowski. ``Quantizing Ho\v{r}ava-Lifshitz gravity via causal dynamical triangulations." \emph{Physical Review D} (2012). 

\bibitem{DB&JH} D. Benedetti and J. Henson.  ``Spectral geometry as a probe of quantum spacetime." \emph{Physical Review D} 80 (2009) 124036.



\bibitem{QGQC} G. Calcagni, L. Papantonopoulos, G. Siopsis, N. Tsamis eds. ``Quantum Gravity and Quantum Cosmology." \emph{Lecture Notes in Physics}. Springer 2013. 


\bibitem{JHC1} J. H. Cooperman. ``Renormalization of lattice-regularized quantum gravity models II. The case of causal dynamical triangulations." arXiv: gr-qc/1406.4531.

\bibitem{JHC&KL&JMM} J. H. Cooperman, K. Lee, and J. M. Miller. ``A closer look at transition amplitudes in $(2+1)$-dimensional causal dynamical triangulations." In preparation.

\bibitem{JHC&JMM} J. H. Cooperman and J. M. Miller. ``A first look at transition amplitudes in $(2+1)$-dimensional causal dynamical triangulations." \emph{Classical and Quantum Gravity} (2014). 

\bibitem{RF&JCH&DNS} R. Flauger, J. C. Hill, and D. N. Spergel. ``Toward an Understanding of Foreground Emission in the BICEP2 Region." \emph{Journal of Cosmology and Astroparticle Physics} 1408 (2014) 039.

\bibitem{SDSS} D. W. Hogg, D. J. Eisenstein, M. R. Blanton, N. A. Bahcall, J. Brinkmann, J. E. Gunn, and D. P. Schneider. ``Cosmic homogeneity demonstrated with luminous red galaxies." \emph{Astrophysics Journal} 624 (2005) 54.

\bibitem{EWK&MT} E. W. Kolb and M. Turner. ``The Early Universe." Westview Press 1994.

\bibitem{RK} R. K. Kommu. ``A validation of causal dynamical triangulations." \emph{Classical and Quantum Gravity} 29 (2012) 105003.

\bibitem{CM&JB&AB} C. Marinoni, J. Bel, and A. Buzzi. ``The scale of cosmic isotropy." \emph{Journal of Cosmology and Astroparticle Physics} 10 (2012) 036.

\bibitem{WiggleZ} M. I. Scrimgeour \emph{et al}. ``The WiggleZ Dark Energy Survey: the transition to large-scale cosmic homogeneity." \emph{Monthly Notices of the Royal Astronomical Society} 425 (2012) 116.


\end{thebibliography}
\end{document}